%% file: scrap.tex
\def\review{0} % With 1 comments are enabled, with 0 not
\def\arxivdisclaimer{1} % With 1, the IEEE disclaimer for arXiv is added

\documentclass[10pt, conference, letterpaper]{IEEEtran}
\usepackage[letterpaper, left = 0.68in, right =0.68in, top = 0.7in, bottom = 1in]{geometry}

\usepackage[noadjust]{cite}
\usepackage{amsmath}
\usepackage{amssymb}
\usepackage{amsthm}
\usepackage{amsfonts}
\usepackage{graphicx}
\usepackage{textcomp}
\usepackage{mathtools}
\usepackage{adjustbox}
\usepackage{balance}
\usepackage{bm}
\usepackage{tikz}
\usetikzlibrary{arrows, fit, matrix, positioning, shapes, backgrounds}
\usepackage{xcolor}
\def\BibTeX{{\rm B\kern-.05em{\sc i\kern-.025em b}\kern-.08em
    T\kern-.1667em\lower.7ex\hbox{E}\kern-.125emX}}

\usepackage{textcase}
\usepackage[tablename=TABLE,font=small]{caption}
\DeclareCaptionTextFormat{up}{\MakeTextUppercase{#1}}
\usepackage{paralist}
\usepackage{supertabular}    
\usepackage{array} 
\usepackage[linesnumbered,ruled]{algorithm2e}
\usepackage{algorithmic}
\usepackage[acronym]{glossaries}

\if\review1
\usepackage{todonotes}
\else
\usepackage[disable]{todonotes}
\fi

\if\arxivdisclaimer0
\else
\usepackage{hyperref}
\glsdisablehyper
\fi
\usepackage{cleveref}

\usepackage{numprint}     
\usepackage{makecell}   
\usepackage{colortbl}
\usepackage{enumitem}            
\usepackage{longtable} 
\usepackage{wrapfig}
\usepackage{booktabs}
\usepackage{graphics}
\usepackage{pgfplots}
\usepgfplotslibrary{groupplots}
\usepackage{pgfplotstable}
\usepackage{subcaption}
\pgfplotsset{compat=1.18} 

\usepackage{tabu}

\newtheorem*{remark*}{Remark}

\usepackage[utf8]{inputenc}

\crefname{figure}{Fig.}{Fig.}
\crefname{table}{Table}{Table}

\let\oldtabular\tabular
\renewcommand{\tabular}{\small\oldtabular}

\newcommand{\egc}{e.\,g., }
\newcommand{\iec}{i.\,e., }

\newcommand{\wrt}{w.\,r.\,t.\ }

\newcolumntype{?}{!{\vrule width 1pt}}
\definecolor{mittelblau}{RGB}{0, 126, 198}
\definecolor{violettblau}{cmyk}{0.9, 0.6, 0, 0}
\definecolor{rot}{RGB}{238, 28 35}
\definecolor{apfelgruen}{RGB}{140, 198, 62}
\definecolor{gelb}{RGB}{1, 221, 0}
\definecolor{orange}{RGB}{244, 111, 33}
\definecolor{pink}{RGB}{237, 0, 140}
\definecolor{lila}{RGB}{128, 10, 145}
\definecolor{hellgrau}{RGB}{224, 224, 224}
\definecolor{mittelgrau}{RGB}{128, 128, 128}
\definecolor{dunkelgrau}{RGB}{80,80,80}
\definecolor{anthrazit}{RGB}{19, 31, 31}
\definecolor{darkgreen}{RGB}{0.125,0.5,0.169}
\definecolor{ahmedyellow}{RGB}{204,153,0}

\newcommand\blfootnote[1]{%
  \begingroup
  \renewcommand\thefootnote{}\footnote{#1}%
  \addtocounter{footnote}{-1}%
  \endgroup
}

\include{./Material/colormap_inue_jetlight}

\begin{document}

\title{CRAP Part II: Clutter Removal with Continuous Acquisitions Under Phase Noise}

\author{
    \IEEEauthorblockN{
        Marcus Henninger\IEEEauthorrefmark{1}\IEEEauthorrefmark{2},
        Silvio Mandelli\IEEEauthorrefmark{1},
        Artjom Grudnitsky\IEEEauthorrefmark{1}, 
        and Stephan ten Brink\IEEEauthorrefmark{2}
        }

	\IEEEauthorblockA{
	\IEEEauthorrefmark{1}Nokia Bell Labs Stuttgart, 70469 Stuttgart, Germany \\
	\IEEEauthorrefmark{2}Institute of Telecommunications, University of Stuttgart, 70569 Stuttgart, Germany \\
	E-mail: marcus.henninger@nokia.com}}
	%,\{firstname.lastname\}@nokia-bell-labs.com}}

\maketitle

\input{Content/Acronyms}
\input{Content/Abstract}

\if\arxivdisclaimer1
\blfootnote{This work has been submitted to the IEEE for possible publication. Copyright may be transferred without notice, after which this version may no longer be accessible.}
\else
\vspace{0.2cm}
\fi

%\fi

\input{Content/Introduction}
\input{Content/System_Model}
\input{Content/Scrap}
\input{Content/Simulation_Results}
\input{Content/Measurement_Results}
\input{Content/Conclusion}
\input{Content/Acknowledgment}

\bibliographystyle{IEEEtran}
\bibliography{scrap}

\end{document}

%% file: Material/colormap_inue_jetlight.tex
\pgfplotsset{
    colormap={jet_inue}{
        rgb=(1, 1, 1)
        rgb=(0.99804, 0.99804, 0.99905)
        rgb=(0.99609, 0.99609, 0.99813)
        rgb=(0.99413, 0.99413, 0.99725)
        rgb=(0.99217, 0.99217, 0.99639)
        rgb=(0.99022, 0.99022, 0.99557)
        rgb=(0.98826, 0.98826, 0.99477)
        rgb=(0.9863, 0.9863, 0.99401)
        rgb=(0.98434, 0.98434, 0.99327)
        rgb=(0.98239, 0.98239, 0.99257)
        rgb=(0.98043, 0.98043, 0.9919)
        rgb=(0.97847, 0.97847, 0.99125)
        rgb=(0.97652, 0.97652, 0.99064)
        rgb=(0.97456, 0.97456, 0.99006)
        rgb=(0.9726, 0.9726, 0.98951)
        rgb=(0.97065, 0.97065, 0.98899)
        rgb=(0.96869, 0.96869, 0.9885)
        rgb=(0.96673, 0.96673, 0.98804)
        rgb=(0.96477, 0.96477, 0.98762)
        rgb=(0.96282, 0.96282, 0.98722)
        rgb=(0.96086, 0.96086, 0.98685)
        rgb=(0.9589, 0.9589, 0.98652)
        rgb=(0.95695, 0.95695, 0.98621)
        rgb=(0.95499, 0.95499, 0.98593)
        rgb=(0.95303, 0.95303, 0.98569)
        rgb=(0.95108, 0.95108, 0.98548)
        rgb=(0.94912, 0.94912, 0.98529)
        rgb=(0.94716, 0.94716, 0.98514)
        rgb=(0.94521, 0.94521, 0.98502)
        rgb=(0.94325, 0.94325, 0.98493)
        rgb=(0.94129, 0.94129, 0.98486)
        rgb=(0.93933, 0.93933, 0.98483)
        rgb=(0.93738, 0.93738, 0.98483)
        rgb=(0.93542, 0.93542, 0.98486)
        rgb=(0.93346, 0.93346, 0.98493)
        rgb=(0.93151, 0.93151, 0.98502)
        rgb=(0.92955, 0.92955, 0.98514)
        rgb=(0.92759, 0.92759, 0.98529)
        rgb=(0.92564, 0.92564, 0.98548)
        rgb=(0.92368, 0.92368, 0.98569)
        rgb=(0.92172, 0.92172, 0.98593)
        rgb=(0.91977, 0.91977, 0.98621)
        rgb=(0.91781, 0.91781, 0.98652)
        rgb=(0.91585, 0.91585, 0.98685)
        rgb=(0.91389, 0.91389, 0.98722)
        rgb=(0.91194, 0.91194, 0.98762)
        rgb=(0.90998, 0.90998, 0.98804)
        rgb=(0.90802, 0.90802, 0.9885)
        rgb=(0.90607, 0.90607, 0.98899)
        rgb=(0.90411, 0.90411, 0.98951)
        rgb=(0.90215, 0.90215, 0.99006)
        rgb=(0.9002, 0.9002, 0.99064)
        rgb=(0.89824, 0.89824, 0.99125)
        rgb=(0.89628, 0.89628, 0.9919)
        rgb=(0.89432, 0.89432, 0.99257)
        rgb=(0.89237, 0.89237, 0.99327)
        rgb=(0.89041, 0.89041, 0.99401)
        rgb=(0.88845, 0.88845, 0.99477)
        rgb=(0.8865, 0.8865, 0.99557)
        rgb=(0.88454, 0.88454, 0.99639)
        rgb=(0.88258, 0.88258, 0.99725)
        rgb=(0.88063, 0.88063, 0.99813)
        rgb=(0.87867, 0.87867, 0.99905)
        rgb=(0.87671, 0.87671, 1)
        rgb=(0.87476, 0.87573, 1)
        rgb=(0.8728, 0.87479, 1)
        rgb=(0.87084, 0.87387, 1)
        rgb=(0.86888, 0.87298, 1)
        rgb=(0.86693, 0.87213, 1)
        rgb=(0.86497, 0.8713, 1)
        rgb=(0.86301, 0.87051, 1)
        rgb=(0.86106, 0.86974, 1)
        rgb=(0.8591, 0.86901, 1)
        rgb=(0.85714, 0.8683, 1)
        rgb=(0.85519, 0.86763, 1)
        rgb=(0.85323, 0.86699, 1)
        rgb=(0.85127, 0.86638, 1)
        rgb=(0.84932, 0.8658, 1)
        rgb=(0.84736, 0.86525, 1)
        rgb=(0.8454, 0.86473, 1)
        rgb=(0.84344, 0.86424, 1)
        rgb=(0.84149, 0.86378, 1)
        rgb=(0.83953, 0.86335, 1)
        rgb=(0.83757, 0.86295, 1)
        rgb=(0.83562, 0.86259, 1)
        rgb=(0.83366, 0.86225, 1)
        rgb=(0.8317, 0.86194, 1)
        rgb=(0.82975, 0.86167, 1)
        rgb=(0.82779, 0.86142, 1)
        rgb=(0.82583, 0.86121, 1)
        rgb=(0.82387, 0.86103, 1)
        rgb=(0.82192, 0.86087, 1)
        rgb=(0.81996, 0.86075, 1)
        rgb=(0.818, 0.86066, 1)
        rgb=(0.81605, 0.8606, 1)
        rgb=(0.81409, 0.86057, 1)
        rgb=(0.81213, 0.86057, 1)
        rgb=(0.81018, 0.8606, 1)
        rgb=(0.80822, 0.86066, 1)
        rgb=(0.80626, 0.86075, 1)
        rgb=(0.80431, 0.86087, 1)
        rgb=(0.80235, 0.86103, 1)
        rgb=(0.80039, 0.86121, 1)
        rgb=(0.79843, 0.86142, 1)
        rgb=(0.79648, 0.86167, 1)
        rgb=(0.79452, 0.86194, 1)
        rgb=(0.79256, 0.86225, 1)
        rgb=(0.79061, 0.86259, 1)
        rgb=(0.78865, 0.86295, 1)
        rgb=(0.78669, 0.86335, 1)
        rgb=(0.78474, 0.86378, 1)
        rgb=(0.78278, 0.86424, 1)
        rgb=(0.78082, 0.86473, 1)
        rgb=(0.77886, 0.86525, 1)
        rgb=(0.77691, 0.8658, 1)
        rgb=(0.77495, 0.86638, 1)
        rgb=(0.77299, 0.86699, 1)
        rgb=(0.77104, 0.86763, 1)
        rgb=(0.76908, 0.8683, 1)
        rgb=(0.76712, 0.86901, 1)
        rgb=(0.76517, 0.86974, 1)
        rgb=(0.76321, 0.87051, 1)
        rgb=(0.76125, 0.8713, 1)
        rgb=(0.7593, 0.87213, 1)
        rgb=(0.75734, 0.87298, 1)
        rgb=(0.75538, 0.87387, 1)
        rgb=(0.75342, 0.87479, 1)
        rgb=(0.75147, 0.87573, 1)
        rgb=(0.74951, 0.87671, 1)
        rgb=(0.74755, 0.87772, 1)
        rgb=(0.7456, 0.87876, 1)
        rgb=(0.74364, 0.87983, 1)
        rgb=(0.74168, 0.88093, 1)
        rgb=(0.73973, 0.88206, 1)
        rgb=(0.73777, 0.88323, 1)
        rgb=(0.73581, 0.88442, 1)
        rgb=(0.73386, 0.88564, 1)
        rgb=(0.7319, 0.88689, 1)
        rgb=(0.72994, 0.88818, 1)
        rgb=(0.72798, 0.88949, 1)
        rgb=(0.72603, 0.89084, 1)
        rgb=(0.72407, 0.89222, 1)
        rgb=(0.72211, 0.89362, 1)
        rgb=(0.72016, 0.89506, 1)
        rgb=(0.7182, 0.89653, 1)
        rgb=(0.71624, 0.89802, 1)
        rgb=(0.71429, 0.89955, 1)
        rgb=(0.71233, 0.90111, 1)
        rgb=(0.71037, 0.9027, 1)
        rgb=(0.70841, 0.90432, 1)
        rgb=(0.70646, 0.90597, 1)
        rgb=(0.7045, 0.90766, 1)
        rgb=(0.70254, 0.90937, 1)
        rgb=(0.70059, 0.91111, 1)
        rgb=(0.69863, 0.91289, 1)
        rgb=(0.69667, 0.91469, 1)
        rgb=(0.69472, 0.91652, 1)
        rgb=(0.69276, 0.91839, 1)
        rgb=(0.6908, 0.92028, 1)
        rgb=(0.68885, 0.92221, 1)
        rgb=(0.68689, 0.92417, 1)
        rgb=(0.68493, 0.92616, 1)
        rgb=(0.68297, 0.92817, 1)
        rgb=(0.68102, 0.93022, 1)
        rgb=(0.67906, 0.9323, 1)
        rgb=(0.6771, 0.93441, 1)
        rgb=(0.67515, 0.93655, 1)
        rgb=(0.67319, 0.93872, 1)
        rgb=(0.67123, 0.94092, 1)
        rgb=(0.66928, 0.94316, 1)
        rgb=(0.66732, 0.94542, 1)
        rgb=(0.66536, 0.94771, 1)
        rgb=(0.66341, 0.95004, 1)
        rgb=(0.66145, 0.95239, 1)
        rgb=(0.65949, 0.95478, 1)
        rgb=(0.65753, 0.95719, 1)
        rgb=(0.65558, 0.95964, 1)
        rgb=(0.65362, 0.96211, 1)
        rgb=(0.65166, 0.96462, 1)
        rgb=(0.64971, 0.96716, 1)
        rgb=(0.64775, 0.96973, 1)
        rgb=(0.64579, 0.97233, 1)
        rgb=(0.64384, 0.97496, 1)
        rgb=(0.64188, 0.97762, 1)
        rgb=(0.63992, 0.98031, 1)
        rgb=(0.63796, 0.98303, 1)
        rgb=(0.63601, 0.98578, 1)
        rgb=(0.63405, 0.98856, 1)
        rgb=(0.63209, 0.99138, 1)
        rgb=(0.63014, 0.99422, 1)
        rgb=(0.62818, 0.9971, 1)
        rgb=(0.62622, 1, 1)
        rgb=(0.6272, 1, 0.99706)
        rgb=(0.62821, 1, 0.9941)
        rgb=(0.62925, 1, 0.9911)
        rgb=(0.63032, 1, 0.98807)
        rgb=(0.63142, 1, 0.98502)
        rgb=(0.63255, 1, 0.98193)
        rgb=(0.63371, 1, 0.97881)
        rgb=(0.63491, 1, 0.97566)
        rgb=(0.63613, 1, 0.97248)
        rgb=(0.63738, 1, 0.96927)
        rgb=(0.63867, 1, 0.96603)
        rgb=(0.63998, 1, 0.96276)
        rgb=(0.64133, 1, 0.95945)
        rgb=(0.6427, 1, 0.95612)
        rgb=(0.64411, 1, 0.95276)
        rgb=(0.64555, 1, 0.94936)
        rgb=(0.64702, 1, 0.94594)
        rgb=(0.64851, 1, 0.94248)
        rgb=(0.65004, 1, 0.939)
        rgb=(0.6516, 1, 0.93548)
        rgb=(0.65319, 1, 0.93193)
        rgb=(0.65481, 1, 0.92836)
        rgb=(0.65646, 1, 0.92475)
        rgb=(0.65815, 1, 0.92111)
        rgb=(0.65986, 1, 0.91744)
        rgb=(0.6616, 1, 0.91374)
        rgb=(0.66337, 1, 0.91001)
        rgb=(0.66518, 1, 0.90625)
        rgb=(0.66701, 1, 0.90246)
        rgb=(0.66888, 1, 0.89864)
        rgb=(0.67077, 1, 0.89478)
        rgb=(0.6727, 1, 0.8909)
        rgb=(0.67466, 1, 0.88699)
        rgb=(0.67665, 1, 0.88304)
        rgb=(0.67866, 1, 0.87907)
        rgb=(0.68071, 1, 0.87506)
        rgb=(0.68279, 1, 0.87102)
        rgb=(0.6849, 1, 0.86696)
        rgb=(0.68704, 1, 0.86286)
        rgb=(0.68921, 1, 0.85873)
        rgb=(0.69141, 1, 0.85457)
        rgb=(0.69365, 1, 0.85039)
        rgb=(0.69591, 1, 0.84617)
        rgb=(0.6982, 1, 0.84192)
        rgb=(0.70053, 1, 0.83763)
        rgb=(0.70288, 1, 0.83332)
        rgb=(0.70527, 1, 0.82898)
        rgb=(0.70768, 1, 0.82461)
        rgb=(0.71013, 1, 0.82021)
        rgb=(0.7126, 1, 0.81577)
        rgb=(0.71511, 1, 0.81131)
        rgb=(0.71765, 1, 0.80681)
        rgb=(0.72022, 1, 0.80229)
        rgb=(0.72282, 1, 0.79773)
        rgb=(0.72545, 1, 0.79314)
        rgb=(0.72811, 1, 0.78853)
        rgb=(0.7308, 1, 0.78388)
        rgb=(0.73352, 1, 0.7792)
        rgb=(0.73627, 1, 0.77449)
        rgb=(0.73905, 1, 0.76975)
        rgb=(0.74187, 1, 0.76498)
        rgb=(0.74471, 1, 0.76018)
        rgb=(0.74758, 1, 0.75535)
        rgb=(0.75049, 1, 0.75049)
        rgb=(0.75342, 1, 0.7456)
        rgb=(0.75639, 1, 0.74067)
        rgb=(0.75939, 1, 0.73572)
        rgb=(0.76241, 1, 0.73074)
        rgb=(0.76547, 1, 0.72572)
        rgb=(0.76856, 1, 0.72068)
        rgb=(0.77168, 1, 0.7156)
        rgb=(0.77483, 1, 0.71049)
        rgb=(0.77801, 1, 0.70536)
        rgb=(0.78122, 1, 0.70019)
        rgb=(0.78446, 1, 0.69499)
        rgb=(0.78773, 1, 0.68976)
        rgb=(0.79103, 1, 0.6845)
        rgb=(0.79437, 1, 0.67921)
        rgb=(0.79773, 1, 0.67389)
        rgb=(0.80113, 1, 0.66854)
        rgb=(0.80455, 1, 0.66316)
        rgb=(0.80801, 1, 0.65775)
        rgb=(0.81149, 1, 0.65231)
        rgb=(0.81501, 1, 0.64683)
        rgb=(0.81855, 1, 0.64133)
        rgb=(0.82213, 1, 0.63579)
        rgb=(0.82574, 1, 0.63023)
        rgb=(0.82938, 1, 0.62463)
        rgb=(0.83305, 1, 0.61901)
        rgb=(0.83675, 1, 0.61335)
        rgb=(0.84048, 1, 0.60766)
        rgb=(0.84424, 1, 0.60194)
        rgb=(0.84803, 1, 0.5962)
        rgb=(0.85185, 1, 0.59042)
        rgb=(0.85571, 1, 0.58461)
        rgb=(0.85959, 1, 0.57877)
        rgb=(0.8635, 1, 0.5729)
        rgb=(0.86745, 1, 0.56699)
        rgb=(0.87142, 1, 0.56106)
        rgb=(0.87543, 1, 0.5551)
        rgb=(0.87946, 1, 0.54911)
        rgb=(0.88353, 1, 0.54308)
        rgb=(0.88763, 1, 0.53703)
        rgb=(0.89176, 1, 0.53094)
        rgb=(0.89591, 1, 0.52483)
        rgb=(0.9001, 1, 0.51868)
        rgb=(0.90432, 1, 0.51251)
        rgb=(0.90857, 1, 0.5063)
        rgb=(0.91285, 1, 0.50006)
        rgb=(0.91717, 1, 0.49379)
        rgb=(0.92151, 1, 0.48749)
        rgb=(0.92588, 1, 0.48116)
        rgb=(0.93028, 1, 0.4748)
        rgb=(0.93472, 1, 0.46841)
        rgb=(0.93918, 1, 0.46199)
        rgb=(0.94368, 1, 0.45554)
        rgb=(0.9482, 1, 0.44906)
        rgb=(0.95276, 1, 0.44255)
        rgb=(0.95734, 1, 0.436)
        rgb=(0.96196, 1, 0.42943)
        rgb=(0.96661, 1, 0.42282)
        rgb=(0.97129, 1, 0.41619)
        rgb=(0.976, 1, 0.40952)
        rgb=(0.98074, 1, 0.40283)
        rgb=(0.98551, 1, 0.3961)
        rgb=(0.99031, 1, 0.38934)
        rgb=(0.99514, 1, 0.38255)
        rgb=(1, 1, 0.37573)
        rgb=(1, 0.99511, 0.37378)
        rgb=(1, 0.99018, 0.37182)
        rgb=(1, 0.98523, 0.36986)
        rgb=(1, 0.98025, 0.36791)
        rgb=(1, 0.97523, 0.36595)
        rgb=(1, 0.97019, 0.36399)
        rgb=(1, 0.96511, 0.36204)
        rgb=(1, 0.96, 0.36008)
        rgb=(1, 0.95487, 0.35812)
        rgb=(1, 0.9497, 0.35616)
        rgb=(1, 0.9445, 0.35421)
        rgb=(1, 0.93927, 0.35225)
        rgb=(1, 0.93401, 0.35029)
        rgb=(1, 0.92872, 0.34834)
        rgb=(1, 0.9234, 0.34638)
        rgb=(1, 0.91805, 0.34442)
        rgb=(1, 0.91267, 0.34247)
        rgb=(1, 0.90726, 0.34051)
        rgb=(1, 0.90182, 0.33855)
        rgb=(1, 0.89634, 0.33659)
        rgb=(1, 0.89084, 0.33464)
        rgb=(1, 0.8853, 0.33268)
        rgb=(1, 0.87974, 0.33072)
        rgb=(1, 0.87414, 0.32877)
        rgb=(1, 0.86852, 0.32681)
        rgb=(1, 0.86286, 0.32485)
        rgb=(1, 0.85717, 0.3229)
        rgb=(1, 0.85146, 0.32094)
        rgb=(1, 0.84571, 0.31898)
        rgb=(1, 0.83993, 0.31703)
        rgb=(1, 0.83412, 0.31507)
        rgb=(1, 0.82828, 0.31311)
        rgb=(1, 0.82241, 0.31115)
        rgb=(1, 0.81651, 0.3092)
        rgb=(1, 0.81057, 0.30724)
        rgb=(1, 0.80461, 0.30528)
        rgb=(1, 0.79862, 0.30333)
        rgb=(1, 0.79259, 0.30137)
        rgb=(1, 0.78654, 0.29941)
        rgb=(1, 0.78045, 0.29746)
        rgb=(1, 0.77434, 0.2955)
        rgb=(1, 0.76819, 0.29354)
        rgb=(1, 0.76202, 0.29159)
        rgb=(1, 0.75581, 0.28963)
        rgb=(1, 0.74957, 0.28767)
        rgb=(1, 0.7433, 0.28571)
        rgb=(1, 0.737, 0.28376)
        rgb=(1, 0.73068, 0.2818)
        rgb=(1, 0.72432, 0.27984)
        rgb=(1, 0.71792, 0.27789)
        rgb=(1, 0.7115, 0.27593)
        rgb=(1, 0.70505, 0.27397)
        rgb=(1, 0.69857, 0.27202)
        rgb=(1, 0.69206, 0.27006)
        rgb=(1, 0.68551, 0.2681)
        rgb=(1, 0.67894, 0.26614)
        rgb=(1, 0.67233, 0.26419)
        rgb=(1, 0.6657, 0.26223)
        rgb=(1, 0.65903, 0.26027)
        rgb=(1, 0.65234, 0.25832)
        rgb=(1, 0.64561, 0.25636)
        rgb=(1, 0.63885, 0.2544)
        rgb=(1, 0.63206, 0.25245)
        rgb=(1, 0.62524, 0.25049)
        rgb=(1, 0.6184, 0.24853)
        rgb=(1, 0.61152, 0.24658)
        rgb=(1, 0.6046, 0.24462)
        rgb=(1, 0.59766, 0.24266)
        rgb=(1, 0.59069, 0.2407)
        rgb=(1, 0.58369, 0.23875)
        rgb=(1, 0.57666, 0.23679)
        rgb=(1, 0.56959, 0.23483)
        rgb=(1, 0.5625, 0.23288)
        rgb=(1, 0.55538, 0.23092)
        rgb=(1, 0.54822, 0.22896)
        rgb=(1, 0.54103, 0.22701)
        rgb=(1, 0.53382, 0.22505)
        rgb=(1, 0.52657, 0.22309)
        rgb=(1, 0.51929, 0.22114)
        rgb=(1, 0.51199, 0.21918)
        rgb=(1, 0.50465, 0.21722)
        rgb=(1, 0.49728, 0.21526)
        rgb=(1, 0.48988, 0.21331)
        rgb=(1, 0.48245, 0.21135)
        rgb=(1, 0.47499, 0.20939)
        rgb=(1, 0.4675, 0.20744)
        rgb=(1, 0.45997, 0.20548)
        rgb=(1, 0.45242, 0.20352)
        rgb=(1, 0.44484, 0.20157)
        rgb=(1, 0.43722, 0.19961)
        rgb=(1, 0.42958, 0.19765)
        rgb=(1, 0.42191, 0.19569)
        rgb=(1, 0.4142, 0.19374)
        rgb=(1, 0.40646, 0.19178)
        rgb=(1, 0.3987, 0.18982)
        rgb=(1, 0.3909, 0.18787)
        rgb=(1, 0.38307, 0.18591)
        rgb=(1, 0.37521, 0.18395)
        rgb=(1, 0.36733, 0.182)
        rgb=(1, 0.35941, 0.18004)
        rgb=(1, 0.35146, 0.17808)
        rgb=(1, 0.34347, 0.17613)
        rgb=(1, 0.33546, 0.17417)
        rgb=(1, 0.32742, 0.17221)
        rgb=(1, 0.31935, 0.17025)
        rgb=(1, 0.31125, 0.1683)
        rgb=(1, 0.30311, 0.16634)
        rgb=(1, 0.29495, 0.16438)
        rgb=(1, 0.28675, 0.16243)
        rgb=(1, 0.27853, 0.16047)
        rgb=(1, 0.27027, 0.15851)
        rgb=(1, 0.26199, 0.15656)
        rgb=(1, 0.25367, 0.1546)
        rgb=(1, 0.24532, 0.15264)
        rgb=(1, 0.23694, 0.15068)
        rgb=(1, 0.22853, 0.14873)
        rgb=(1, 0.2201, 0.14677)
        rgb=(1, 0.21163, 0.14481)
        rgb=(1, 0.20312, 0.14286)
        rgb=(1, 0.19459, 0.1409)
        rgb=(1, 0.18603, 0.13894)
        rgb=(1, 0.17744, 0.13699)
        rgb=(1, 0.16882, 0.13503)
        rgb=(1, 0.16016, 0.13307)
        rgb=(1, 0.15148, 0.13112)
        rgb=(1, 0.14277, 0.12916)
        rgb=(1, 0.13402, 0.1272)
        rgb=(1, 0.12524, 0.12524)
        rgb=(0.99315, 0.12329, 0.12329)
        rgb=(0.98627, 0.12133, 0.12133)
        rgb=(0.97936, 0.11937, 0.11937)
        rgb=(0.97242, 0.11742, 0.11742)
        rgb=(0.96545, 0.11546, 0.11546)
        rgb=(0.95845, 0.1135, 0.1135)
        rgb=(0.95141, 0.11155, 0.11155)
        rgb=(0.94435, 0.10959, 0.10959)
        rgb=(0.93726, 0.10763, 0.10763)
        rgb=(0.93013, 0.10568, 0.10568)
        rgb=(0.92298, 0.10372, 0.10372)
        rgb=(0.91579, 0.10176, 0.10176)
        rgb=(0.90857, 0.099804, 0.099804)
        rgb=(0.90133, 0.097847, 0.097847)
        rgb=(0.89405, 0.09589, 0.09589)
        rgb=(0.88674, 0.093933, 0.093933)
        rgb=(0.8794, 0.091977, 0.091977)
        rgb=(0.87203, 0.09002, 0.09002)
        rgb=(0.86463, 0.088063, 0.088063)
        rgb=(0.8572, 0.086106, 0.086106)
        rgb=(0.84974, 0.084149, 0.084149)
        rgb=(0.84225, 0.082192, 0.082192)
        rgb=(0.83473, 0.080235, 0.080235)
        rgb=(0.82718, 0.078278, 0.078278)
        rgb=(0.81959, 0.076321, 0.076321)
        rgb=(0.81198, 0.074364, 0.074364)
        rgb=(0.80434, 0.072407, 0.072407)
        rgb=(0.79666, 0.07045, 0.07045)
        rgb=(0.78896, 0.068493, 0.068493)
        rgb=(0.78122, 0.066536, 0.066536)
        rgb=(0.77345, 0.064579, 0.064579)
        rgb=(0.76566, 0.062622, 0.062622)
        rgb=(0.75783, 0.060665, 0.060665)
        rgb=(0.74997, 0.058708, 0.058708)
        rgb=(0.74208, 0.056751, 0.056751)
        rgb=(0.73416, 0.054795, 0.054795)
        rgb=(0.72621, 0.052838, 0.052838)
        rgb=(0.71823, 0.050881, 0.050881)
        rgb=(0.71022, 0.048924, 0.048924)
        rgb=(0.70218, 0.046967, 0.046967)
        rgb=(0.6941, 0.04501, 0.04501)
        rgb=(0.686, 0.043053, 0.043053)
        rgb=(0.67787, 0.041096, 0.041096)
        rgb=(0.6697, 0.039139, 0.039139)
        rgb=(0.66151, 0.037182, 0.037182)
        rgb=(0.65328, 0.035225, 0.035225)
        rgb=(0.64503, 0.033268, 0.033268)
        rgb=(0.63674, 0.031311, 0.031311)
        rgb=(0.62842, 0.029354, 0.029354)
        rgb=(0.62008, 0.027397, 0.027397)
        rgb=(0.6117, 0.02544, 0.02544)
        rgb=(0.60329, 0.023483, 0.023483)
        rgb=(0.59485, 0.021526, 0.021526)
        rgb=(0.58638, 0.019569, 0.019569)
        rgb=(0.57788, 0.017613, 0.017613)
        rgb=(0.56935, 0.015656, 0.015656)
        rgb=(0.56079, 0.013699, 0.013699)
        rgb=(0.5522, 0.011742, 0.011742)
        rgb=(0.54357, 0.0097847, 0.0097847)
        rgb=(0.53492, 0.0078278, 0.0078278)
        rgb=(0.52624, 0.0058708, 0.0058708)
        rgb=(0.51752, 0.0039139, 0.0039139)
        rgb=(0.50878, 0.0019569, 0.0019569)
        rgb=(0.5, 0, 0)
    }
}

%% file: Content/Acronyms.tex
\newacronym{3GPP}{3GPP}{3rd Generation Partnership Project}
\newacronym{5G}{5G}{fifth generation}
\newacronym{6G}{6G}{sixth generation}
\newacronym{awgn}{AWGN}{additive white Gaussian noise}
\newacronym{cfar}{CFAR}{constant false alarm rate}
\newacronym{crap}{CRAP}{Clutter Removal with Acquisitions Under Phase Noise}
\newacronym{csi}{CSI}{channel state information}
\newacronym{dl}{DL}{downlink}
\newacronym{eca}{ECA}{Extensive Cancellation Algorithm}
\newacronym{eca-c}{ECA-C}{Extensive Cancellation Algorithm by Subcarrier}
\newacronym{eca-s}{ECA-S}{Extensive Cancellation Algorithm by Symbol}
\newacronym{dft}{DFT}{Discrete Fourier Transform}
\newacronym{idft}{IDFT}{Inverse Discrete Fourier Transform}
\newacronym{gnb}{gNB}{gNodeB}
\newacronym{isac}{ISAC}{Integrated Sensing and Communication}
\newacronym{kf}{KF}{Kalman Filter}
\newacronym{mdl}{MDL}{Minimum Description Length}
\newacronym{mp}{MP}{Marchenko-Pastur}
\newacronym{ofdm}{OFDM}{orthogonal frequency-division multiplexing}
\newacronym{poc}{PoC}{proof of concept}
\newacronym{rmse}{RMSE}{root-mean-square error}
\newacronym{rf}{RF}{radio frequency}
\newacronym{rx}{RX}{receiver}
\newacronym{scnr}{SCNR}{signal-to-clutter-noise ratio}
\newacronym{scrap}{SCRAP}{Smoothed CRAP}
\newacronym{svd}{SVD}{singular value decomposition}
\newacronym{snr}{SNR}{signal-to-noise ratio}
\newacronym{tdd}{TDD}{Time Division Duplex}
\newacronym{tx}{TX}{transmitter}

%% file: Content/Abstract.tex
%\todo[inline]{MH: title to be discussed
%\newline SM: see proposal. It takes the old one and enhances it with a bit (remember that an average reader did not read the previous). Btw, I like more ``allows to cope'', rather than ``enables coping'', since ``enable/enabler'' is more of a standalone \textit{keyword}, but I leave it up to you
%\newline MH: ok, I would have omitted the phase noise since it's almost not addressed in this paper, but I don't mind having it there, so let's keep it.
%\newline regarding the safety thing: to me it's not so clear why ``safety in factory floors'' is a scenario that is not completely static. To me this comes a bit out of the blue.
%\newline SM: I see. Do you have a way to make the abstract a bit more grounded in reality/use cases? We can have a quick chat if needed.
%\newline MH: see new suggestion.}

\begin{abstract}

The mitigation of clutter is an important research branch in \gls{isac}, one of the emerging technologies of future cellular networks. In this work, we extend our previously introduced method \gls{crap} by means to track clutter over time. This is necessary in scenarios that require high reliability but can change dynamically, like safety applications in factory floors. To that end, exponential smoothing is leveraged to process new measurements and previous clutter information in a unique matrix using the singular value decomposition, allowing adaptation to changing environments in an efficient way. We further propose a singular value threshold based on the Marchenko-Pastur distribution to select the meaningful clutter components. Results from both simulations and measurements show that continuously updating the clutter components with new acquisitions according to our proposed algorithm \gls{scrap} enables coping with dynamic clutter environments and facilitates the detection of sensing targets.

\end{abstract}

%% file: Content/Introduction.tex
\glsresetall

\section{Introduction}\label{sec:intro}
%\todo[inline]{MH: Maybe cut \cite{wild2021joint} at the end, if we need more space}

\gls{isac} is envisioned to augment cellular networks with radar-like capabilities~\cite{de2021convergent}, ideally with close to zero overhead compared to legacy communications services~\cite{wild2021joint}. The feasibility of \gls{isac} for beyond \gls{5G} and \gls{6G} cellular networks is currently being assessed both in standardization discussions within the framework of the~\gls{3GPP}~\cite{3gpp_22837} as well as by the research community, \egc through analytical studies~\cite{mandelli2023survey, liu2022survey}.

One of the various \gls{isac} research challenges is the mitigation of clutter, \iec of reflections caused by objects that are not of interest for the sensing task and can thus be regarded as interference. In a previous work~\cite{henninger2023crap}, we introduced \gls{crap}, which builds on the principles of the~\gls{eca}~\cite{colone2009multistage} and is tailored to the requirements in \gls{isac} deployments. By leveraging vectorization and the \gls{svd}, the approach exploits the multi-dimensionality of the \gls{csi}, thereby enabling the reliable detection of slow or non-moving targets in contrast to prior art~\cite{zhao2012multipath, liu2019evaluation}. Further, the algorithm can cope with phase incoherence, \egc due to \gls{tx} and \gls{rx} not sharing the same local oscillator as in~\cite{wild2023integrated}. 

% \todo[inline]{SM: rephrased into ``single round of measurements acquisition'' to better signal that we take multiple measures at once (before one could suspect it was a single measure). Btw, CRAP does not give a **** on RCS, but it's rather, range, speed, and/or shape, right? (since we project, it does not care how strong is the projection, it's anyway removed) However, to determine the meaningful clutter components, RCS is still important. If you agree - at least on the content - please rephrase.
% Btw, Shall we add a ref to the ARENA's website?
% \newline MH: - agree on multiple measures thing, but slightly rephrased since ``single round of measurements acquisition'' sounds a bit cumbersome to me. 
% \newline - good point regrading the RCS; for now I simply removed it, if there's still space at the end I'll spend a sentence or two on it.
% \newline - regarding ARENA website ref: added it for now, we also did it in the positioning paper. However, as I don't think it adds too much and takes away some space, it will be the first thing to be removed if we need more space :)}

\begin{figure}
    \centering
    \input{Figures/ranges_over_time.tikz}
    \caption
    {Estimated range of the strongest clutter component over a time span of ten seconds for two different measurements recorded within half an hour (time stamps in legend).}
    \label{fig:ranges_over_time}
\end{figure}
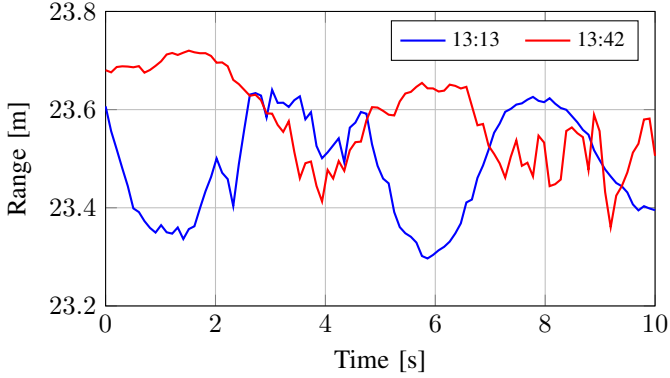

However, as \gls{crap} determines the clutter ``offline'' in a single round of measurements, it also inherently relies on the clutter components being constant, \iec not changing their properties (\egc range) over time, which does not always hold true in practice. For instance, clutter may exhibit large-scale variations that are difficult to capture in a single acquisition round. Moreover, clutter components can simply \mbox{(dis-)appear} with time. 
Fig.~\ref{fig:ranges_over_time} shows a real-world example of a dynamic clutter environment based on measurements from the \gls{isac} \gls{poc}~\cite{wild2023integrated} taken in the ARENA2036 industrial research campus. The figure plots the estimated range of the strongest clutter component over ten seconds for two different measurements recorded within half an hour. One can discern that the range does not only exhibit strong fluctuations within the respective ten-second intervals, but also shows different behavior when comparing the two measurements. Such circumstances naturally also cause the initially obtained clutter information to deviate from reality with time, which impacts how well the clutter is removed at runtime and can ultimately impair the sensing performance. Further, in practice it can not always be assumed to know when the environment is target-free –- or that it ever becomes target-free for a long enough period in the first place. The initial round of acquisitions used for determining the clutter components in \gls{crap}~\cite{henninger2023crap} may therefore also contain target contributions.

%\footnote{https://www.arena2036.de/en/} 

The above reasons call for a method that allows to dynamically update the clutter information with new measurements. Essentially, the clutter components, \iec their \gls{csi} contributions, need to be tracked to account for changes and uncertainties in the environment. Such an algorithm should clearly also be computationally lightweight for real-time execution. 

In principle, various prior art approaches can be used for that, such as the well-known \gls{kf}~\cite{kalman1960new}, which is commonly employed in radar applications~\cite{ramachandra2018kalman}. However, tracking multiple components/objects with a \gls{kf} (or related methods) typically incurs additional problems to be solved, like data association between measurements, requiring supplementary algorithms with potentially prohibitive complexity. In this work, we make use of exponential smoothing~\cite{brown1961fundamental}, which is utilized in various fields of application, \egc for time-series forecasting in economics~\cite{de2011forecasting}. %or in signal processing to deal with noisy data, as done, \egc in~\cite{henninger2023performance} to prevent cycle slips. 
% \todo[inline]{SM: a bit stretched the reference to~\cite{henninger2023performance}. In case we violate the 15-20\% rule, let's swap it with another one please.
% \newline Below I also added a sentence on the clutter removal step, feel free to delete it if you do not like it, but I thought it would be good to recall that tracking is not the only thing that must be done.
% \newline MH: good point, I removed the reference; initial motivation was to also have a reference its use in engineering/signal processing, not only in economics. Let me know if you have maybe another suitable reference from the top of your head, otherwise we leave it like that.
% \newline Sentence about clutter removal is fine thanks :)}

The solution discussed in this paper, \gls{scrap}, extends \gls{crap}'s clutter acquisition step by combining the most recent measurements with the previously acquired clutter information. After updating the clutter information, clutter removal is performed as in~\cite{henninger2023crap}. The main contributions of this work can be summarized as follows:

% \todo[inline]{MH: bullet points w/ main contributions to be improved/extended if space allows.
% \newline SM: it's ok, but remember that the contribution is ``an extension to the clutter acquisition procedure in~\cite{henninger2023crap} to cope with...'', and then you might explain how/what you do, rather than a simple ``we track ...'' where you just focus on what you do
% \newline MH: good point, added a half sentence at the beginning of the firs bullet point (hope that's what you mean)
% \newline SM: the ``we track'' formulation is a bit passive. I would start the contribution with something a bit more active, like my proposal in this comment or ``We design a method to track ...'' or similar}

\begin{itemize}
    \item To cope with dynamic clutter environments, we design a method to track the clutter by constantly updating the clutter subspace with new measurements using exponential smoothing. To do this efficiently, the previous clutter components are scaled by their singular values, enabling stacking them with the new measures in a single matrix.
    \item A threshold based on the \gls{mp} distribution is proposed, which allows to select the meaningful clutter components after applying the \gls{svd}.
    \item We validate \gls{scrap} both with simulations and measurements, showing that our proposed algorithm can adapt to dynamic clutter environments and thereby improves the clutter removal capability.
\end{itemize}

%% file: Figures/ranges_over_time.tikz
\begin{tikzpicture}
    \centering
    \begin{axis}[
        height=5.5cm,
	width=0.49\textwidth,
        grid=major, 
        xlabel={Time [s]},
    	ylabel={Range [m]},
		xlabel near ticks, 
        ylabel near ticks,
        xmin = 0,
        xmax = 10,
        ymin = 23.2,
        ymax = 23.8,
		legend cell align={left},
		legend style={/tikz/every even column/.append style={column sep=2.5mm}},
	 	legend columns=2,
		legend pos = north east,
        legend style={font=\footnotesize}
		]
            
        \addplot
        [blue, 
        thick] 
        plot table[x index=0, y index=1] {Data/ranges_3a.txt}; 

        \addlegendentry{13:13}

        \addplot[red, 
        thick] 
        plot table[x index=0, y index=1] {Data/ranges_4.txt}; 

        \addlegendentry{13:42}

	\end{axis}
\end{tikzpicture}

%% file: Content/System_Model.tex
\section{System Model}\label{sec:sys_model}

We briefly recall the system model used in~\cite{henninger2023crap} with co-located, but physically separated \gls{tx} and \gls{rx}. The \gls{tx} transmits $M$ \gls{ofdm} symbols over $N$ subcarriers spaced by $\Delta f$ at carrier frequency $f_c$, denoted as the transmitted frame $\mathbf{X} \in \mathbb{C}^{N \times M}$ carrying phase-modulated constant envelope complex symbols.

Each object in the surroundings indexed $p~\in~\mathcal{P}$ with range~$r_p$ and velocity $v_p$ relative to the \gls{rx} generates a reflected path. The objects are partitioned into sets consisting of clutter components and sensing targets, denoted by $\mathcal{P}_c \subseteq \mathcal{P}$ and $\mathcal{P}_t \subseteq \mathcal{P}$, respectively.
The received frame $\mathbf{Y} \in \mathbb{C}^{N \times M}$ can be expressed as a superposition of the reflected paths caused by the objects in the environment. Removing the influence of the transmitted symbols $\mathbf{X}$ at the \gls{rx} via element-wise division of $\mathbf{Y}$ by $\mathbf{X}$, the time-frequency \gls{csi} matrix writes as
%\todo[inline]{TW: of $\mathbf{Y}$ by $\mathbf{X}$} 
\begin{align}
\mathbf{H} = \phi \sum_{p \in \mathcal{P}} \alpha_{p}\mathbf{a}(r_p)\mathbf{b}(v_p)^\text{T} + \textbf{Z} \;,
\label{eq:Y}	
\end{align} 
where $\mathit{\alpha_{p}}$ is the complex coefficient of the $\mathit{p}$-th path and $\phi$ the random phase rotation. Further, $\mathbf{Z} \in \mathbb{C}^{N \times M}$ represents the random complex \gls{awgn} matrix, where each element has a power of $\sigma_n^2 = P_n/N$, with $P_n$ being the noise power over the whole bandwidth.
% \todo[inline]{SM: some peaky comments: AWGN with constant variance holds only if the received noise was AWGN in the first place, and all elements in $\mathbf{X}$ were with constant modulus (e.g. QPSK symbols).
% \newline MH: good point. Should I add something like ``phase-modulated constant envelope
% complex symbols'' (as in the 2D MUSIC paper)? However, then we slightly diverge from the formulation of the system model in the initial CRAP paper, which I'm not sure if we want to do.
% \newline SM: Whatever works for you, works for me, but it's important to mention these assumptions somehow
% \newline MH: ok, please check. To me this suggests also a bit that we restrict ourselves to constant amplitude symbols, but you are right, of course.}
The vectors $\mathbf{a}(\mathit{r_{p}})$ and $\mathbf{b}(v_p)$ describing the channel contributions due to range $r_p$ and velocity $v_p$ of the $p$-th object are
\begin{align}
\mathbf{a}(r_{p}) &= \begin{bmatrix}
       1, \ e^{-j4\pi \Delta f \cdot r_p/c}, \ \dots, \ e^{-j4\pi (N - 1) \Delta f \cdot r_p/c}
\end{bmatrix}^\text{T} \\
\mathbf{b}(v_p) &= \begin{bmatrix}
       1, \ e^{j4\pi T_0 f_c \cdot v_p / c}, \ \dots, \ e^{j4\pi (M - 1) T_0  f_c \cdot v_p / c}
\end{bmatrix}^\text{T}  \; , 
\label{eq:channel_vectors}
\end{align}
where $c$ denotes the speed of light and $T_0$ the duration of an \gls{ofdm} symbol. 

Even though the system model assumes a mono-static setup, the following approach can also be applied to bi-static or multi-static deployments. For more details, \egc about the reasoning behind the phase noise modeling, please refer to~\cite{henninger2023crap}. 

%% file: Content/Scrap.tex
\section{Dynamic Clutter Removal}\label{sec:scrap}

As \gls{scrap} extends \gls{crap}, we briefly recall its principles in the first part of this section. Then, the features of \gls{scrap} that allow adapting to dynamic clutter environments are discussed. 

\subsection{CRAP Recap}\label{subsec:crap}

\subsubsection{Clutter Acquisition (Offline)}\label{subsubsec:crap_acquisition}

Initially, at time $t=0$, $K_0$~snapshots of the reference scenario are recorded. To exploit the multi-dimensionality of the \gls{csi} in contrast to prior art~\cite{zhao2012multipath}, the $K_0$ vectorized \gls{csi} frames~\eqref{eq:Y} with $Q=MN$ elements are stacked along the rows to form the clutter acquisition matrix $\mathbf{C}_0~\in~\mathbb{C}^{K_0 \times Q}$. Next, the compact \gls{svd} of $\mathbf{C}_0$ is performed
\begin{align}
\mathbf{C}_0 = \mathbf{U}_0 \mathbf{\Sigma}_0 \mathbf{V}_0^{\text{H}} \; ,
\label{eq:svd}
\end{align}
where the columns of $\mathbf{U}_0\in~\mathbb{C}^{K_0 \times K_0}$ and rows of $\mathbf{V}_0^{\text{H}} \in~\mathbb{C}^{K_0 \times Q}$ are the left and right singular vectors, respectively, of $\mathbf{C}_0$, and $(\cdot)^{\text{H}}$ denotes the Hermitian transpose. In order to determine the number of meaningful clutter components $L_0$, or -- hereafter -- the ``clutter order'', based on the singular values on the diagonal of $\mathbf{\Sigma}_0 \in~\mathbb{R}^{K_0 \times K_0}$,  Minimum Description Length~\cite{rissanen1978modeling} is used. The right singular vectors (rows) of $\mathbf{V}_0^{\text{H}}$ associated with the~$L_0$ strongest singular values are stacked along the columns and form the clutter subspace $\hat{\mathbf{C}}_0~\in~\mathbb{C}^{Q \times L_0}$. % comprising the most significant clutter components.

To make clutter removal at runtime computationally feasible, part of the projection matrix is pre-computed as 
\begin{align}
\mathbf{P}_0^{\prime} = \hat{\mathbf{C}}_0\bigl(\hat{\mathbf{C}}_0^{\text{H}}\hat{\mathbf{C}}_0\bigr)^{-1} \; 
\label{eq:clutter_projection}
\end{align}
and stored along with the Hermitian transpose of the clutter subspace $\hat{\mathbf{C}}_0^{\text{H}}~\in~\mathbb{C}^{L_0 \times Q}$.

\subsubsection{Clutter Removal (Runtime)}

At runtime, the clutter is removed from the vectorized sensing acquisition $\mathbf{h}$ by subtracting its projection into the clutter subspace as% using $\mathbf{P}^{\prime}$ and $\hat{\mathbf{C}}^{\text{H}}$ as
\begin{align}
\hat{\mathbf{h}} &= \mathbf{h} - \mathbf{P}_0^{\prime}\bigl(\hat{\mathbf{C}}_0^{\text{H}}\mathbf{h}\bigr)  \; .
\label{eq:clutter_removal}
\end{align}
The clutter-rejected \gls{csi} matrix $\hat{\mathbf{H}}$ used for \gls{ofdm} radar processing is obtained by reshaping $\hat{\mathbf{h}}$ into its original shape.

For a more detailed discussion about the advantages of the previously outlined two-step approach, please refer to \cite{henninger2023crap}.

\subsection{Dynamic Clutter Acquisition: SCRAP}\label{subsec:scrap}

\gls{crap} is now extended to enable tracking of clutter information by continuously updating the clutter subspace with new measurements during nominal sensing operations. 
To that end, exponential smoothing~\cite{brown1961fundamental} is leveraged to write the \textit{smoothed} clutter matrix at time index~$t$ as 
\begin{align}
\mathbf{C}_t^{\prime} = 
 \begin{cases}
   \qquad \sqrt{K_t}^{-1} \mathbf{C}_0  & t = 0 \\
   \begin{bmatrix}
	\rho_t \sqrt{K_t}^{-1} \mathbf{C}_t \\
	\sqrt{1-\rho_t^2} \left( \hat{\mathbf{C}}_{t-1}^{\prime} \hat{\mathbf{\Sigma}}_{t-1}^{\prime} \right)^{\text{T}}
	\end{bmatrix}		& t > 0	
   \end{cases} \; ,
   \label{eq:smoothed_clutter_mat}
\end{align}
%\todo[inline]{SM: you did not introduce $\mathbf{C}_t$, did you? You are indirectly doing it in the next sentence, but it's a bit misleading compared to $\mathbf{C}$. How about introducing it also in CRAP recap, and then using always $\mathbf{C}_0$ there?
%\newline Please, if possible add an equation (even inline if no space) with $\mathbf{C}' = \mathbf{\Sigma}'\mathbf{C}$, everything with the hat, to help a bit the reader to follow (if you have space)
%\newline MH: good point, added $\mathbf{C}_0$ (and, accordingly, also the $0$ subscript everywhere else to be consistent), please have a look.
%\newline Introducing $\mathbf{C}' = \mathbf{\Sigma}'\mathbf{C}$ would conflict with already existing notation, no? Should I still introduce it, \egc with a new variable? Suggestions?
%\newline SM: ok to leave it like this to avoid excessive notation. Feel free to comment this out.}
where for $t>0$ the current clutter acquisition matrix $\mathbf{C}_t \in~\mathbb{C}^{K_t \times Q}$ is merged with the smoothed clutter subspace of the previous iteration $\hat{\mathbf{C}}_{t-1}^{\prime} \in~\mathbb{C}^{Q \times L_{t-1}}$ via exponential smoothing with factor $\rho_t$. To achieve that, the components from the previous clutter subspace are multiplied by their singular values $\hat{\mathbf{\Sigma}}_{t-1}^{\prime} \in~\mathbb{R}^{L_{t-1} \times L_{t-1}}$. This scales the $L_{t-1}$ components in $\hat{\mathbf{C}}_{t-1}^{\prime}$ (unit vectors) to their relevance (in terms of how much energy they scatter back) in previous measurements and allows direct merging with $\mathbf{C}_t$ by stacking the two matrices in a single matrix $\mathbf{C}_t^{\prime} \in~\mathbb{C}^{(K_t + L_{t-1}) \times Q}$~\eqref{eq:smoothed_clutter_mat}. %as done in~\eqref{eq:smoothed_clutter_mat}. 

Here, it is crucial to emphasize that multiplying with the corresponding left singular vectors $\hat{\mathbf{U}}_{t-1}^{\prime} \in~\mathbb{C}^{(K_{t-1} + L_{t-1}) \times L_{t-1}} $ is \textit{not} required and would even entail computational overhead. This is due to the fact that the relevance of each measurement contributing to the generation of the subspace is not of interest, but only the extracted subspace components. Therefore, our approach only requires $\hat{\mathbf{C}}_{t-1}^{\prime} \hat{\mathbf{\Sigma}}_{t-1}^{\prime} \in~\mathbb{C}^{Q \times L_{t-1}}$ from the previous update, which can be stored as a single matrix. As it is an exponential moving average of the clutter components, clutter information of previous measurements is inherently incorporated in a matrix with only $L_{t-1}$ columns, enabling efficient tracking of the significant clutter components.
% \todo[inline]{SM: Very good explanations
% \newline MH: thanks, they are partly yours (e.g., the remarks about the left singular vectors)}

Note that in~\eqref{eq:smoothed_clutter_mat} the clutter acquisition matrix $\mathbf{C}_t$ is normalized via multiplication with $\sqrt{K_t}^{-1}$  to ensure that the correct scaling between updates in case of varying numbers of acquisitions~$K_t$ is preserved. In the initial clutter acquisition step ($t=0$), there is obviously no previous clutter information available, so that the first clutter matrix $\mathbf{C}_0^{\prime}$ is obtained as described in~\ref{subsec:crap} (plus additional multiplication with $\sqrt{K_t}^{-1}$). 

% \todo[inline]{MH: @Silvio: ``energy relationship'' term ok?
% \newline SM: how about ``right/correct scaling''?
% \newline MH: I like it, thanks}

\subsection{Selecting Clutter Components}

%\todo[inline]{MH: motivate necessity of MP threshold a bit more (also benefits compared to model order estimation techniques)
%\newline SM: shall I do it? If not, you have to introduce before the ``For that, ...'' below, that you are setting up a statistical test with null hyphotesis being that there is no clutter, but just AWGN, and you want to test it against the hyphothesis that there is something else in a certain bin bla bla... Then you can start explaining that you consider just an AWGN matrix bla bla...
%\newline MH: yes, if you like to do it, I'd be happy :) I fear it will take up quite some space though ...
%\newline SM: have a look please
%\newline MH: great, thanks, I like it. With ``necessity'' I also meant to write a bit why we moved away from model order techniques, but I guess it's ok to not mention it due to space reasons.}

\todo[inline]{MH: add comment about why MP threshold is necessary and model order is not good enough (if space allows)}
%After obtaining $\mathbf{C}_t^{\prime}$, the meaningful clutter components must again be extracted, %for which the \gls{svd} is first applied as per~\eqref{eq:svd}. However, we now propose to determine the clutter order with a threshold based on the \gls{mp} distribution~\cite{marchenko1967distribution}. 
After obtaining the clutter matrix $\mathbf{C}_t^{\prime}$, the meaningful clutter components must be determined. For this purpose, we design a statistical test with the null hypothesis corresponding to the case where only Gaussian noise $\mathbf{Z}$ from~\eqref{eq:Y} has generated $\mathbf{C}_t^{\prime}$. Under this hypothesis, the clutter acquisition matrix $\mathbf{C}_t$ would have entries that are standard complex Gaussian random variables with variance corresponding to the noise power $\sigma_n^2$. 

It has been shown that the distribution of the eigenvalues of the sample correlation matrix $\frac{1}{Q}\mathbf{C}_t\mathbf{C}_t^\text{H}$ is then known to approximately follow the \gls{mp} distribution~\cite{marchenko1967distribution}. However, the \gls{mp} distribution typically approximates the eigenvalue distribution well overall, but in our studies we observed that it is not tight at the  tails of the distribution. For this reason, in order to design a test with low probability of false alarms, we do not determine the threshold of the statistical test  by selecting a quantile value of the \gls{mp} distribution, but rather by considering the derived upper bound for the eigenvalues of the sample correlation matrix~\cite{marchenko1967distribution} 
\begin{align}
\lambda_{\text{max}} = \sigma_n^2 \left(1+\sqrt{\frac{K_t}{Q}} \right)^2 \; .
\label{eq:ev_mp_upper_bound}
\end{align}
% \todo[inline]{SM: since it's out/your contribution, I'd suggest to write ``We propose to determine the singular value threshold to count as valid one component and, hence, to determine $L_t$, by using (8) and by ...'' You can even split the two sentences, but do not skip important explanations and to highlight your contribution.
% \newline MH: ok, pls check}
Finally, we propose to determine that a singular value -- as square root of the sample correlation matrix eigenvalue -- was generated by a relevant clutter component if it exceeds the threshold
\begin{align}
\eta_{\text{SV} \!, \, t} = 
 \begin{cases}
   \sigma_n \left(1+\sqrt{\frac{Q}{K_t}} \right)  & t = 0 \\
   \rho_t \sigma_n \left(1+\sqrt{\frac{Q}{K_t}} \right)  & t > 0	
   \end{cases} \; ,
   \label{eq:sv_threshold_scrap}
\end{align}
where the scaling with $\sqrt{K_t}^{-1}$ and $\rho_t$ (for $t > 0$), and the missing division with $Q$ are accounted for to normalize the values according to~\eqref{eq:smoothed_clutter_mat}. The number of singular values exceeding $\eta_{\text{SV} \!, \, t}$ then determines the clutter order $L_t$.

% $\lambda_{\text{max}}$ of~\eqref{eq:ev_mp_upper_bound}, 
% $L_t$. The threshold is derived from~\eqref{eq:ev_mp_upper_bound} by factoring in the scaling with $\sqrt{K_t}^{-1}$ and the missing division with $Q$ (for the sample correlation matrix) as
% \begin{align}
% \eta_{\text{SV} \!, \, t} = 
%  \begin{cases}
%    \sigma_n \left(1+\sqrt{\frac{Q}{K_t}} \right)  & t = 0 \\
%    \rho_t \sigma_n \left(1+\sqrt{\frac{Q}{K_t}} \right)  & t > 0	
%    \end{cases} \; ,
%    \label{eq:sv_threshold_scrap}
% \end{align}
% where for $t > 0$ the multiplication with the smoothing parameter $\rho_t$  in~\eqref{eq:smoothed_clutter_mat} is taken into account. 

% \todo[inline]{MH: @Silvio: next paragraph (``leaves some degrees [...]'' etc.) was added in the thesis to explain that threshold is not 100 \% straightforward anymore compared to the case where we only have a clutter acquisition matrix (as in CRAP), therefore it's also a bit of a heuristic. Ok to write it like that?
% \newline SM: see my new text (yours commented out). I would not go deep, but still hint that it's a heuristic and that we considered other alternatives
% \newline MH: thanks :) I like it (but slightly rephrased)}

Note that the fact that $\mathbf{C}_t^{\prime}$ from~\eqref{eq:smoothed_clutter_mat} is a \textit{composition} of the current clutter acquisition matrix and the previous clutter components makes what is described above a heuristic method instead of a formal technique.
We anyway left the formal derivation assuming white Gaussian noise for the null hypothesis here, since we thought it would help the reader understand how we started our reasoning. 
Also other alternatives were considered and investigated in our simulation studies, such as scaling the threshold with $\rho_t$ for $t>0$. However, for space reasons and due to the fact that the heuristic method was observed to be the most robust one from every considered aspect, in this paper we limit our description and analysis to the threshold as per~\eqref{eq:sv_threshold_scrap}. 

% The fact that $\mathbf{C}_t^{\prime}$ is a \textit{composition} of the current clutter acquisition matrix and the previous clutter components leaves some degrees of freedom for the threshold derivation. 
%Recall for comparison $\eta_{\text{SV}}$ from Eq.~\eqref{eq:sv_threshold}, which can be straightforwardly derived, as the singular values are generated by a \textit{single} clutter acquisition matrix.
% Scaling the threshold with $\rho_t$ (for $t > 0$) has the effect that components are added to the clutter subspace irrespective of the smoothing parameter. However, this does \textit{not} render the choice of $\rho_t$ obsolete, since it still gives control about how long components remain in the clutter subspace. On the other hand, omitting the scaling with $\rho_t$, \iec choosing the same $\eta_{\text{SV}}$ for all smoothing parameters, requires contributions to be stronger before being added to the clutter subspace. Using this threshold lead to worse results in experiments and is not further considered.

Finally, based on the clutter order $L_t$, the clutter subspace $\hat{\mathbf{C}}_t^{\prime} \in~\mathbb{C}^{Q \times L_t}$ and the clutter removal matrices are obtained as described in~\ref{subsubsec:crap_acquisition}. Accordingly, clutter removal at runtime is also performed in the same way with~\eqref{eq:clutter_removal}. \gls{scrap} can therefore be seen as an extension of \gls{crap}'s clutter acquisition step, which allows tracking clutter components over time via continuous updates with new measurements.

%\todo[inline] {MH: algo box? probably not enough space :(
%\newline SM: Probably not, but I'd add the references to the CRAP steps equations, at least the ones you have in this work}

%% file: Content/Simulation_Results.tex
\section{Simulation Results}\label{sec:sim_results}

% \todo[inline]{SM: not sure we need these two lines here. A generic intro, or - better - nothing would also do. Btw, do you want to say that - in contrast to~\cite{henninger2023crap} - we are in a different Scenario in the next Section (you do not mention it)?

% Btw, how about recalling the system model equation, saying that you generate acquisitions according to (1)?
% \newline MH: ok, removed. I typically like to add some text before a subsection, but I guess it's ok to go without it. Good point regarding linking to the system model.}

%\todo[inline]{SM: I miss a simulation parameters table (if I miss it, sorry)? (or a brief recall of the most important ones in the text, then referring to~\cite{henninger2023crap})}

%This section demonstrates the benefits of updating the clutter components with \gls{scrap} in a dynamic clutter environment.

\subsection{Simulation Setup}

The simulation setup is based on the system model from Section~\ref{sec:sys_model} and the scenario from~\cite{henninger2023crap}, \iec the same $\lvert \mathcal{P}_c \rvert = 5$ randomly generated clutter components are present and the \gls{rf} parameters from Table~\ref{tab:clutter_rf_params} are used. However, the clutter environment now changes over time to investigate the benefits of continuously updating the clutter subspace with measurements using~\gls{scrap}. For that, one clutter component is modeled to change its range with both a sinusoidal component as well as a linear drift away from its initial position at roughly 11.5~m from the sensing system. This results in a displacement of its range of ca. 0.5~m over the entire simulation time, which is in the order of magnitude that can also be observed in Fig.~\ref{fig:ranges_over_time}. 

\begin{table}
    \caption{RF simulation parameters. \label{tab:clutter_rf_params}}
	\centering
     \begin{tabu}{|l|r|}
            \hline
            \textbf{Parameter} & \textbf{Value} \\
			\Xhline{3\arrayrulewidth}
			Carrier frequency $f_c$ & 27.4 GHz \\
  			\hline
  			Number of subcarriers $N$ & 1584 \\
  			\hline
  			Subcarrier spacing $\Delta f$ & 120 kHz \\
  			\hline
            Total bandwidth $B$ & 190 MHz \\
  			\hline
  			Number of \gls{ofdm} symbols per radio frame $M$ & 1120 \\
  			\hline
    \end{tabu}
\end{table}

The single target to be detected is not generated randomly for each trial as in \cite{henninger2023crap}, but moving back and forth with sinusoidal velocity (max. $2  \; \frac{\text{m}}{\text{s}}$) between $5$~m and $20$~m from the system. Such a continuous movement is required to investigate how updating the clutter subspace with measurements containing target contributions affects the performance.

% \todo[inline]{SM: we already commented on the target-free initial scenario, and this red text sentence seems to be in contrast with it. Thus, either we refine the story better also in the intro/abstract, or we do not comment on this assumption at all (I would go for it also to save space)
% \newline MH: ok, I don't see how it is in contrast with it tbh, but I removed it. Good to save space :)}
As in~\cite{henninger2023crap}, $K_0=100$ radio frames are initially recorded to determine the clutter components according to the \gls{crap} principles (\ref{subsubsec:crap_acquisition}). In this initial clutter acquisition, the scenario is target-free. Each radio frame is used for sensing, resulting in an update rate of 0.01~s. The simulation campaign comprises 10000 sensing acquisitions, amounting to an overall time span of 100~s. The clutter subspace is updated every 10~s using the~10 most recent sensing acquisitions, \iec $K_t = 10, \; t > 0$. 

We compare \gls{scrap} with smoothing parameters $\rho = \{0.25, \, 0.5, \,0.75, \, 1\}$ to \gls{crap}, which only uses the clutter components from the initial clutter acquisition for clutter removal, \iec without updates.
% \todo[inline]{SM: how about saying that we consider fixed $rho_t = \rho$ hereafter somewhere?
% \newline MH: good point, done.}
The smoothing parameter is constant for all setups, \iec $\rho_t = \rho$.
For $\rho = 1$, \gls{scrap} coincides with fully recomputing the clutter subspace every $10$~s according to the first step of \gls{crap}, \iec only the 10 most recent acquisitions are considered without tracking. Further, we restrict the maximum number of components allowed in the clutter subspace to 10. This heuristic avoids a performance degradation for low noise powers, where target contributions are more likely to be added to the clutter subspace.
% \todo[inline]{SM: we could (not necessary) add a comment on this hand-tuned parameter, saying that it would be easy to tune according to the specific scenario

% Btw, we don't need Martin Braun to just recall the periodogram concept (just use CRAP) 
% \newline MH: ok, removed (it's anyway already cited earlier, so it would not even have created a new reference entry).
% \newline SM: thanks. I also removed MB reference above (it was at the element-wise division statement, thus not necessary)
% \newline MH: ok :)}

The different setups are investigated \wrt probability of missed detection $P_{\text{MD}}$ and \gls{scnr}. To detect the sensing target, we only process the strongest peak. For more details on the target detection based on the periodogram, please refer to our previous work~\cite{henninger2023crap}.

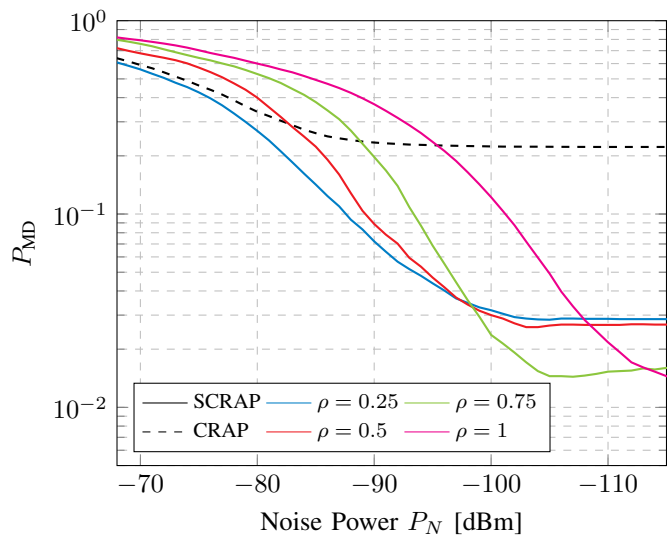
\begin{figure}[!t]
\input{Figures/missed_det_prob_changing.tikz} 
  \caption{Probability of missed detection for the baseline \glsentryshort{crap} (dashed) and \glsentryshort{scrap} (solid) with smoothing parameters $\rho = \{0.25, \, 0.5, \,0.75, \, 1\}$ marked by the corresponding colors.} 
  \label{fig:prob_det}
\end{figure}
%\todo[inline]{SM: we can probably remove $\rho = 0.1$ from the figures, it anyway overlaps with $\rho = 0.25$
%\newline MH: change}

\subsection{Detection Performance}

The $P_{\text{MD}}$ curves in Fig.~\ref{fig:prob_det} reveal that continuously updating the clutter subspace with new measurements brings advantages, as the \gls{scrap} setups overall outperform the \gls{crap} baseline, especially at low noise powers. With a properly chosen smoothing parameter $\rho$, \gls{scrap} reaches a missed detection probability of almost 0.01, whereas \gls{crap} already saturates at roughly 0.2. Choosing a small smoothing parameter appears beneficial for this scenario, as $\rho = 0.25$ exhibits a lower $P_{\text{MD}}$ than \gls{crap} over the whole noise power range. For very low noise powers, trusting the most recent measurements more is advantageous, as $\rho = 0.75$ and $\rho = 1$ achieve the lowest $P_\text{MD}$ floors. Still, as a consequence of the environment being more challenging due to the non-static clutter component, more missed detections occur overall compared to the results in the static clutter environment discussed in~\cite{henninger2023crap}.
% \todo[inline]{SM: do you want to comment on the lower MD floors with extremely low noise, when we (more/just) trust the current measurements?
% \newline MH: good point, thanks}

\subsection{SCNR Performance}

Next, the different setups are compared \wrt the \gls{scnr}
\begin{align}
\gamma_c = \frac{\text{PER}\left(\breve{n}, \breve{m} \right)}{P_c} \; ,
\label{eq:peak_scnr}
\end{align}
where $\breve{n}$ and $\breve{m}$ denote the indices of the periodogram bin in which the target peak is expected according to the ground truth range and velocity. The residual clutter power $P_c$ is computed by averaging those periodogram bins that lie within ellipses with height and width corresponding to five times the range and velocity resolution, respectively, and centered at the ground truth range and velocity bins of the clutter components. 
% \todo[inline]{SM: did you introduce $\Delta r/f$? In case not, we can write what they are.
% \newline MH: thanks, copy pasted from thesis, where it has been defined earlier. Here not, so it's explained now.}

Compared to $P_{\text{MD}}$, this metric offers the advantage that it is solely based on the periodogram, \iec it does not depend on the chosen peak detection strategy. The \gls{scnr} provides a measure of how well the setups isolate the target \wrt clutter, which is typically correlated with the achievable target detection performance, irrespective of how peaks are processed.

\begin{figure}[!t]
  \centering
  \input{Figures/scnr_changing.tikz}
\caption{\glsentryshort{scnr} for the baseline \glsentryshort{crap} (dashed) and \glsentryshort{scrap} (solid) with smoothing parameters $\rho = \{0.25, \, 0.5, \,0.75, \, 1\}$ marked by the corresponding colors.} 
  \label{fig:scnr_changing}
\end{figure}
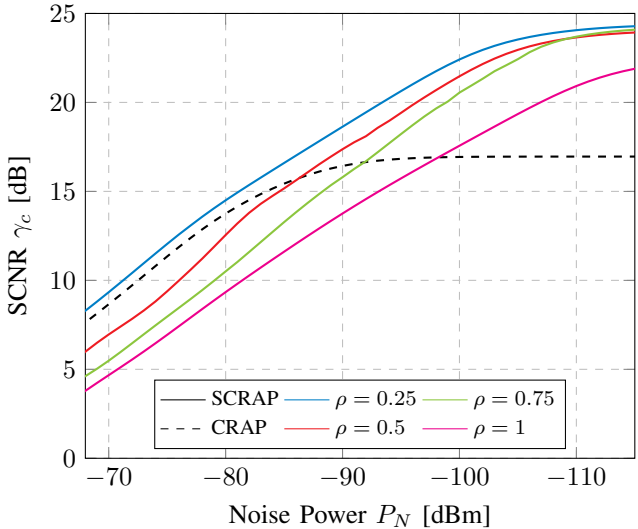

The \gls{scnr} curves in Fig.~\ref{fig:scnr_changing} show that \gls{scrap} with $\rho = 0.25$ again performs best, providing the highest \gls{scnr} and outperforming \gls{crap} over the whole noise power range. For low $P_N$, a gain of ca. $7$~dB over the baseline is attained.

\subsection{Remarks on Smoothing Parameter}
%\todo[inline]{SM: In case of lack of space this can be strongly compressed and added after subsection B.}

Some remarks about the smoothing parameter~$\rho$ are in order. In our scenario, choosing $\rho$ small appears to be advantageous. However, this should at best be regarded as a rough guideline rather than a general rule of thumb, as the optimal $\rho$ depends on the clutter environment and is thus application-specific. 

A small $\rho$ has the effect that components, once added to the clutter subspace, take longer to fall below the threshold $\eta_{\text{SV}}^{\prime}$. This can be unfavorable under certain conditions, since then also erroneously added components, \egc contributions caused by targets, remain in the clutter subspace longer. Moreover, in experiments in an even more dynamic clutter environment we observed that selecting a higher $\rho$ has proven to be beneficial. 

In that context, it should be noted that our simulation setup makes assumptions about clutter behavior and movement of the sensing target. The results thus reflect how suitable a certain $\rho$ is \textit{for this specific scenario}. Other setups, \egc practical applications, might present different challenges, requiring also to tailor the parameters accordingly. Still, the main takeaway from the study in the changing clutter environment is that regularly updating the clutter subspace with new measurements is beneficial -- even in the presence of sensing targets.

%The remarks at the end of \ref{sec:scrap} (and possibly other aspects) should therefore be taken into account when choosing the smoothing parameter in practice. 

%\begin{figure*}[!t]
%\centering
%\begin{subfigure}{0.49\textwidth}
   %\input{Figures/range_rmse_changing.tikz} 
   %\caption{Range RMSE.}
  % \label{fig:range_rmse}
%\end{subfigure}
%begin{subfigure}{0.49\textwidth}
   %\input{Figures/speed_rmse_changing.tikz} 
   %\caption{Velocity RMSE.}
   %\label{fig:speed_rmse}
%\end{subfigure}

%\caption{Range and velocity \glsentryshort{rmse} for the baseline \glsentryshort{crap} (dashed curve) and \glsentryshort{scrap} (solid curves) with different smoothing parameters $\rho = \{0.1, \, 0.25, \, 0.5, \,0.75, \, 1\}$ marked by the corresponding colors.}
%\label{fig:rmse}
%\end{figure*}

%\subsection{RMSE Performance}

%A similar trend is also discernible for the range and velocity \gls{rmse} curves displayed in Fig.~\ref{fig:rmse}. Continuously updating the clutter subspace with \gls{scrap} also results in an improved range and velocity estimation capability compared to \gls{crap}. In particular, the best \gls{scrap} setups with $\rho = \{0.1, \, 0.25\}$ -- which again outperform \gls{crap} over the whole noise power range -- achieve gains of roughly $0.15$~m and $0.08 \; \frac{\text{m}}{\text{s}}$ \wrt range and velocity estimation error, respectively, over the baseline for low noise powers. In those high \gls{snr} regimes, a larger $\rho$ leads to even higher gains, with $\rho = \{0.75, \, 1\}$ exhibiting the best performance.

%% file: Figures/missed_det_prob_changing.tikz
%\tikzsetnextfilename{scrap_prob_det}

\def\scale{1}

\begin{tikzpicture}
		\begin{semilogyaxis}[
			height = 7.5cm,
    		width=0.49\textwidth,
    		xlabel={Noise Power $P_N$ [dBm]},
    		ylabel={$P_{\text{MD}}$},
    		xmax = -68,
            xmin = -115,
            x dir=reverse,
    		ymin=0.005,
		    ymax = 1,
            %ylabel style={font=\footnotesize,at={(axis description cs:.-0.05,.5)},rotate=0,anchor=south},
            %ylabel shift = -14 pt,
		    enlargelimits = false,
    		xmajorgrids=true,
    		yminorgrids=true,
    		grid style=dashed,
    		legend pos = south west,
		    legend style={font=\footnotesize},
            legend columns = 2,
            transpose legend,
            legend cell align={left},
			every axis plot/.append style={thick},
        	scale = \scale,
		]
    	  	
    	\addplot[
   		color=black,
   		style=dashed,
        mark options={solid},
        forget plot]
    	plot table[x expr=\thisrowno{0}, y index=1] {Data/Missed_Det_CRAP_Alpha_Threshold_1_Max_Components_10.txt};
         
        %\addplot[
   	%color=ahmedyellow,
        %mark options={solid},
        %forget plot]
    	%plot table[x expr=\thisrowno{0}, y index=1] {Data/Missed_Det_SCRAP_Alpha_Threshold_1_Max_Components_10_Alpha_0.1.txt};
        
        \addplot[
   		color=mittelblau,
        mark options={solid},
        forget plot]
    	plot table[x expr=\thisrowno{0}, y index=1] {Data/Missed_Det_SCRAP_Alpha_Threshold_1_Max_Components_10_Alpha_0.25.txt};

        \addplot[
   		color=rot,
        mark options={solid},
        forget plot]
    	plot table[x expr=\thisrowno{0}, y index=1] {Data/Missed_Det_SCRAP_Alpha_Threshold_1_Max_Components_10_Alpha_0.5.txt};

        \addplot[
   		color=apfelgruen,
        mark options={solid},
        forget plot]
    	plot table[x expr=\thisrowno{0}, y index=1] {Data/Missed_Det_SCRAP_Alpha_Threshold_1_Max_Components_10_Alpha_0.75.txt};

        \addplot[
   		color=pink,
        mark options={solid},
        forget plot]
    	plot table[x expr=\thisrowno{0}, y index=1] {Data/Missed_Det_SCRAP_Alpha_Threshold_1_Max_Components_10_Alpha_1.txt};

        \addplot[thin, color=black, solid, draw=none] coordinates {(0, 100)}; \addlegendentry{\glsentryshort{scrap}}
		\addplot[thin, color=black, dashed, draw=none] coordinates {(0, 100)}; \addlegendentry{\glsentryshort{crap}}
        %\addplot[thin, color=ahmedyellow, solid, draw=none] coordinates {(0, 100)}; \addlegendentry{$\rho = 0.1$}
		\addplot[thin, color=mittelblau, solid, draw=none] coordinates {(0, 100)}; \addlegendentry{$\rho = 0.25$}
		\addplot[thin, color=rot, solid, draw=none] coordinates {(0, 100)}; \addlegendentry{$\rho = 0.5$}
		\addplot[thin, color=apfelgruen, solid, draw=none] coordinates {(0, 100)};  \addlegendentry{$\rho = 0.75$}
		\addplot[thin, color=pink, solid, draw=none] coordinates {(0, 100)}; \addlegendentry{$\rho = 1$}
        
	\end{semilogyaxis}
\end{tikzpicture}

%% file: Figures/scnr_changing.tikz
%\tikzsetnextfilename{scrap_snr}

\def\scale{1}

\begin{tikzpicture}
		\begin{axis}[
			height = 7.5cm,
    		width=0.49\textwidth,
    		xlabel={Noise Power $P_N$ [dBm]},
    		ylabel={SCNR $\gamma_c$ [dB]},
    		xmax = -68,
            xmin = -115,
            x dir=reverse,
    		ymin= 0,
		    ymax = 25,
		    enlargelimits = false,
    		xmajorgrids=true,
    		ymajorgrids=true,
    		grid style=dashed,
            legend style={
                at={(0.5, 0.01)},
                anchor=south,
                font=\footnotesize,
                inner sep=2pt,
                outer sep=2pt},
            legend columns = 2,
            transpose legend,
            legend cell align={left},
			every axis plot/.append style={thick},
        	scale = \scale,
		]
    	  	
    	\addplot[
   		color=black,
   		style=dashed,
        mark options={solid},
        forget plot]
    	plot table[x expr=\thisrowno{0}, y index=1] {Data/SNR_CRAP_Alpha_Threshold_1_Max_Components_10.txt};
         
        %\addplot[
   		%color=ahmedyellow,
        %mark options={solid},
        %forget plot]
    	%plot table[x expr=\thisrowno{0}, y index=1] {Data/SNR_SCRAP_Alpha_Threshold_1_Max_Components_10_Alpha_0.1.txt};

        \addplot[
   		color=mittelblau,
        mark options={solid},
        forget plot]
    	plot table[x expr=\thisrowno{0}, y index=1] {Data/SNR_SCRAP_Alpha_Threshold_1_Max_Components_10_Alpha_0.25.txt};

        \addplot[
   		color=rot,
        mark options={solid},
        forget plot]
    	plot table[x expr=\thisrowno{0}, y index=1] {Data/SNR_SCRAP_Alpha_Threshold_1_Max_Components_10_Alpha_0.5.txt};

        \addplot[
   		color=apfelgruen,
        mark options={solid},
        forget plot]
    	plot table[x expr=\thisrowno{0}, y index=1] {Data/SNR_SCRAP_Alpha_Threshold_1_Max_Components_10_Alpha_0.75.txt};

        \addplot[
   		color=pink,
        mark options={solid},
        forget plot]
    	plot table[x expr=\thisrowno{0}, y index=1] {Data/SNR_SCRAP_Alpha_Threshold_1_Max_Components_10_Alpha_1.txt};

        \addplot[thin, color=black, solid, draw=none] coordinates {(0, 100)}; \addlegendentry{\glsentryshort{scrap}}
		\addplot[thin, color=black, dashed, draw=none] coordinates {(0, 100)}; \addlegendentry{\glsentryshort{crap}}
        %\addplot[thin, color=ahmedyellow, solid, draw=none] coordinates {(0, 100)}; \addlegendentry{$\rho = 0.1$}
		\addplot[thin, color=mittelblau, solid, draw=none] coordinates {(0, 100)}; \addlegendentry{$\rho = 0.25$}
		\addplot[thin, color=rot, solid, draw=none] coordinates {(0, 100)}; \addlegendentry{$\rho = 0.5$}
		\addplot[thin, color=apfelgruen, solid, draw=none] coordinates {(0, 100)};  \addlegendentry{$\rho = 0.75$}
		\addplot[thin, color=pink, solid, draw=none] coordinates {(0, 100)}; \addlegendentry{$\rho = 1$}
        
	\end{axis}
\end{tikzpicture}

%% file: Content/Measurement_Results.tex
\section{Measurement Results}\label{sec:meas_results}

\begin{figure}[!t]
\centering
  \includegraphics
  [width=7.165cm,
  height = 7.165cm]
  {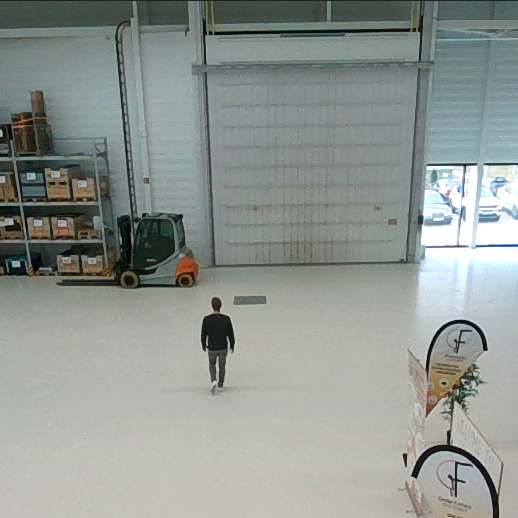}
\caption{Measurement scenario with dynamic clutter component (cargo gate with plastic curtain) and human sensing target from the perspective of the sensing system.}
\label{fig:cargo_gate_target}
\end{figure}

Finally, we validate \gls{scrap} with measurements conducted in the ARENA2036 using the \gls{isac} \gls{poc} described in~\cite{wild2023integrated}. Fig.~\ref{fig:cargo_gate_target} shows the underlying scenario with the strong clutter component from Fig.~\ref{fig:ranges_over_time} (cargo gate with plastic curtain) and a moving pedestrian from the perspective of the sensing system. To account for the dynamic behavior of the clutter, we update the clutter subspace every 5 seconds using the last 10 sensing frames according to the principles described in \ref{subsec:scrap}.

\begin{figure*}[!t]
  \centering
  \hspace{-3mm}
  \subfloat
  	 [Clutter removal with \glsentryshort{crap}. 
  		\label{fig:crap_after_1}]
  		{
		\input{Figures/crap_after_update_1.tikz}
		} 
		\hspace{-4mm}
  \subfloat
  		[Clutter removal with \glsentryshort{scrap} after first update.
  		\label{fig:scrap_after_1}]
  		{
		\input{Figures/scrap_after_update_1.tikz}
		} 
		\\[2mm]
  \subfloat
  		[Clutter removal with \glsentryshort{scrap} before second update.
  		\label{fig:scrap_before_2}]
  		{
		\input{Figures/scrap_before_update_2.tikz}
		} 
		\hspace{-4mm}
  \subfloat
  		[Clutter removal with \glsentryshort{scrap} after second update.
  		\label{fig:scrap_after_2}]
  		{
		\input{Figures/scrap_after_update_2.tikz}
		} 
		%\quad
  \caption{Periodograms after clutter removal with (S)CRAP. The periodograms above are from a frame just after a clutter update, showing that \gls{scrap} (Fig.~\ref{fig:scrap_after_1}) removes the clutter better than \gls{crap}~(Fig.~\ref{fig:crap_after_1}). Fig.~\ref{fig:scrap_before_2} shows a periodogram after \gls{scrap} from a frame just before the next update, where higher clutter contributions are observed. After the next update with \gls{scrap} (Fig.~\ref{fig:scrap_after_2}), the  clutter power is lower again.}  
\label{fig:periodograms_measurements}
\vspace{-4.1mm}
\end{figure*}
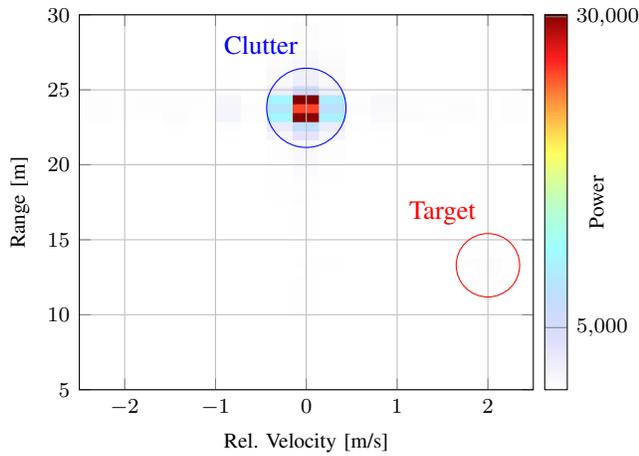
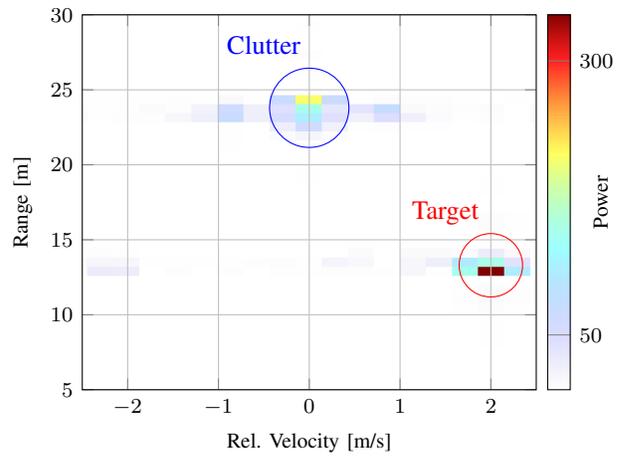
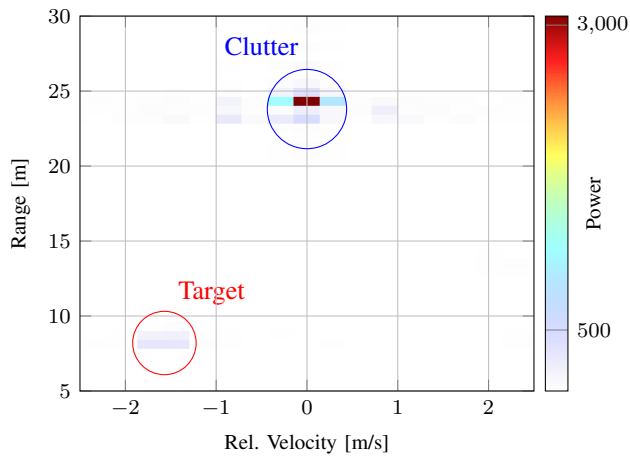
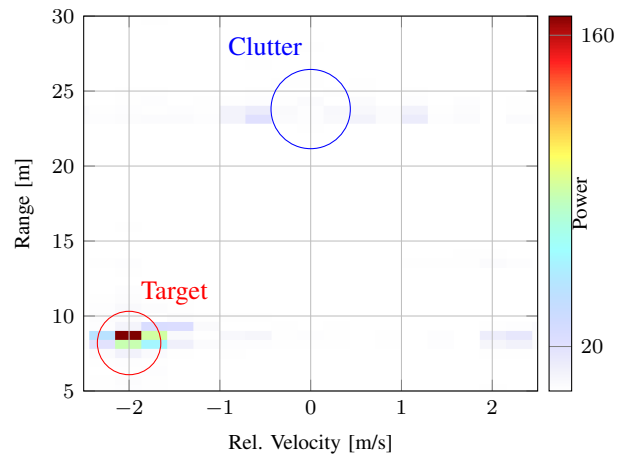

Fig.~\ref{fig:periodograms_measurements} 
depicts periodograms with \gls{crap} (\ref{fig:crap_after_1}), where the components were captured using $K_0=100$ frames from a target-free reference scenario, and \gls{scrap}~(\ref{fig:scrap_after_1} -- \ref{fig:scrap_after_2}), where the clutter information initially acquired as with \gls{crap} is updated every 5 seconds according to our proposal (\ref{subsec:scrap}) using the 10 most recent sensing acquisitions and $\rho = 0.5$. The periodograms in the top row are from the same sensing acquisition from a point in time directly after a clutter update. It can be seen that adapting to the dynamic environment by updating the clutter information with \gls{scrap}~(\ref{fig:scrap_after_1}) reduces the residual clutter considerably compared to \gls{crap}~(\ref{fig:crap_after_1}), which removes the clutter only based on the initial clutter information. With \gls{scrap}, the target is the strongest contribution in the periodogram. The bottom periodograms are from acquisitions just before and right after another update, respectively, \iec ca. 5 seconds later. One can discern that after the update~(\ref{fig:scrap_after_2}), the residual clutter is significantly weaker than before~(\ref{fig:scrap_before_2}). This observation motivates frequent updates.

\todo[inline]{MH: classify results a bit more (short measurement duration, very specific scenario etc.) either here or in conclusion.
\newline SM: naaaah, just say here that this is a measurement trace example. Then you can say in the conclusion that a more holistic evaluation of the complete system will be object of future work.}

%\todo[inline]{SM: I don't like 2 CRAP figures. You can have 4 figures, CRAP at time 1 (1), SCRAP at time 1 before update, SCRAP at time 2 before update, SCRAP at time 2 after update. In this way, you can compare CRAP with SCRAP directly (1-2), showing degradation just before an update (3), and improvement again (4). Or you can test different $\rho$. However, I leave it up to you if this takes too much time
%\newline MH: great comment :) see new figures (4) according to your proposal. Please have a look and let me know if this is what you meant (still have to refine the text describing the results a bit).
%\newline SM: Really nice figures. Now I would simply say ``after first SCRAP update'' ``just before second SCRAP update'', ``after second SCRAP update'', or something similar (don't like much time 1, 2, 3)
%\newline MH: I understand, I also don't like time 1, 2, 3 much. Reason I put it is to show that the CRAP periodogram (top left) is from the same acquisition as SCRAP (top right). If I use ``after first SCRAP update'' also there, it suggests that also CRAP is updated. Do you know what I mean? Would you still remove it?}

\balance

%% file: Figures/crap_after_update_1.tikz
\pgfplotsset{scaled y ticks=false}

\begin{tikzpicture}

    \begin{axis}[
        axis on top,
        width=0.42\textwidth,
        %scale only axis,
        %axis equal=true,
        grid=major,
        enlargelimits=false,
        xmin=-2.5, xmax=2.5,
        xtick={-2, ..., 2},
        xlabel near ticks,
        ymin=5, ymax=30,
        ytick={0,5,...,30},
        ylabel near ticks,
        xlabel={Rel. Velocity  [m/s]},
        ylabel={Range [m]},
        label style={font=\footnotesize},
        tick label style={font=\footnotesize},
        legend style={font=\footnotesize},
        legend pos = north east,
        legend style=
        	{fill=white, 
        	fill opacity=0.4, 
        	draw opacity=1, 
        	text opacity=1, 
        	nodes={scale=1, transform shape}, 
            /tikz/every even column/.append style={column sep=0.1cm}
        	},
        %colorbar horizontal,
        colorbar,
        point meta min=0,
        point meta max=30158.47,
        colormap name=jet_inue,
        colorbar style=
            {ylabel={Power}, 
            at={(1.05, 0)},
            anchor = south,
            width = 0.3cm,
            ytick={5000, 30000},
            ylabel style={yshift=0.9cm},
            every axis/.append style=
                {font=\footnotesize}
            }
        ]

      \addplot[forget plot] graphics[xmin=-3.3, xmax=3, ymin=0, ymax=30] {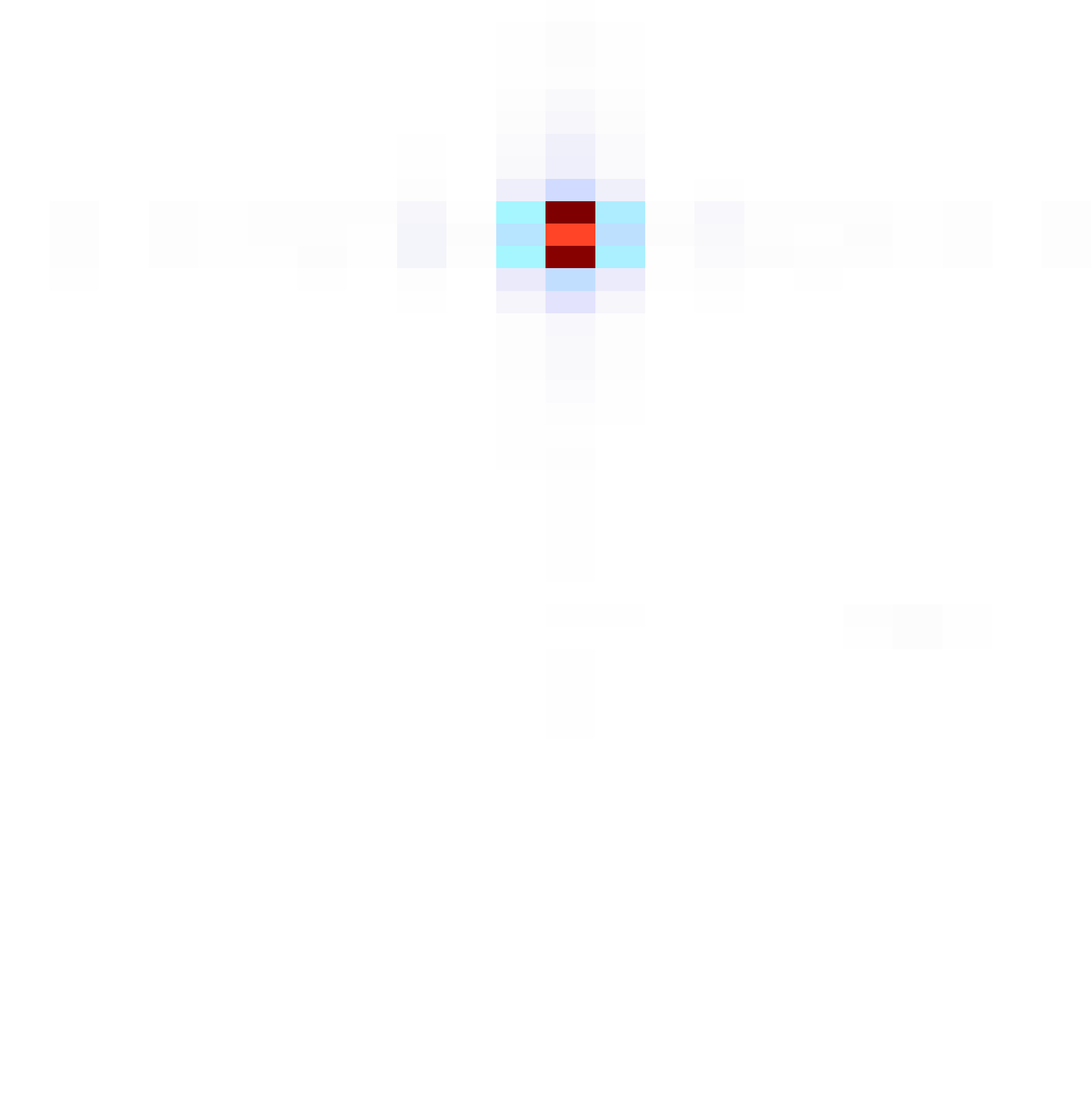};

      	\addplot[
      		mark=o,
                only marks,
      		mark size=15pt,
      		color=blue] 
      		table[x=speed,y=range] {
                speed range
                0 23.8
                };

      	\addplot[
      		mark=o,
      		only marks,
      		mark size=12pt,
      		color=red] 
      		table[x=speed,y=range] {
                speed range
                2 13.3
                };
            
        \node[draw=none, color=blue] at (-0.5, 28) {Clutter};
        \node[draw=none, color=red] at (1.5, 16.8) {Target};

    \end{axis}

\end{tikzpicture}

%% file: Figures/scrap_after_update_1.tikz
\begin{tikzpicture}

    \begin{axis}[
        axis on top,
        width=0.42\textwidth,
        %scale only axis,
        %axis equal=true,
        grid=major,
        enlargelimits=false,
        xmin=-2.5, xmax=2.5,
        xtick={-2, ..., 2},
        xlabel near ticks,
        ymin=5, ymax=30,
        ytick={0,5,...,30},
        ylabel near ticks,
        xlabel={Rel. Velocity  [m/s]},
        ylabel={Range [m]},
        label style={font=\footnotesize},
        tick label style={font=\footnotesize},
        legend style={font=\footnotesize},
        legend pos = north east,
        legend style=
        	{fill=white, 
        	fill opacity=0.4, 
        	draw opacity=1, 
        	text opacity=1, 
        	nodes={scale=1, transform shape}, 
            /tikz/every even column/.append style={column sep=0.1cm}
        	},
        %colorbar horizontal,
        colorbar,
        point meta min=0,
        point meta max=342.04,
        colormap name=jet_inue,
        colorbar style=
            {ylabel={Power}, 
            at={(1.05, 0)},
            anchor = south,
            width = 0.3cm,
            ytick={50, 300},
            ylabel style={yshift=0.5cm},
            every axis/.append style=
                {font=\footnotesize}
            }
        ]

      \addplot[forget plot] graphics[xmin=-3.3, xmax=3, ymin=0, ymax=30] {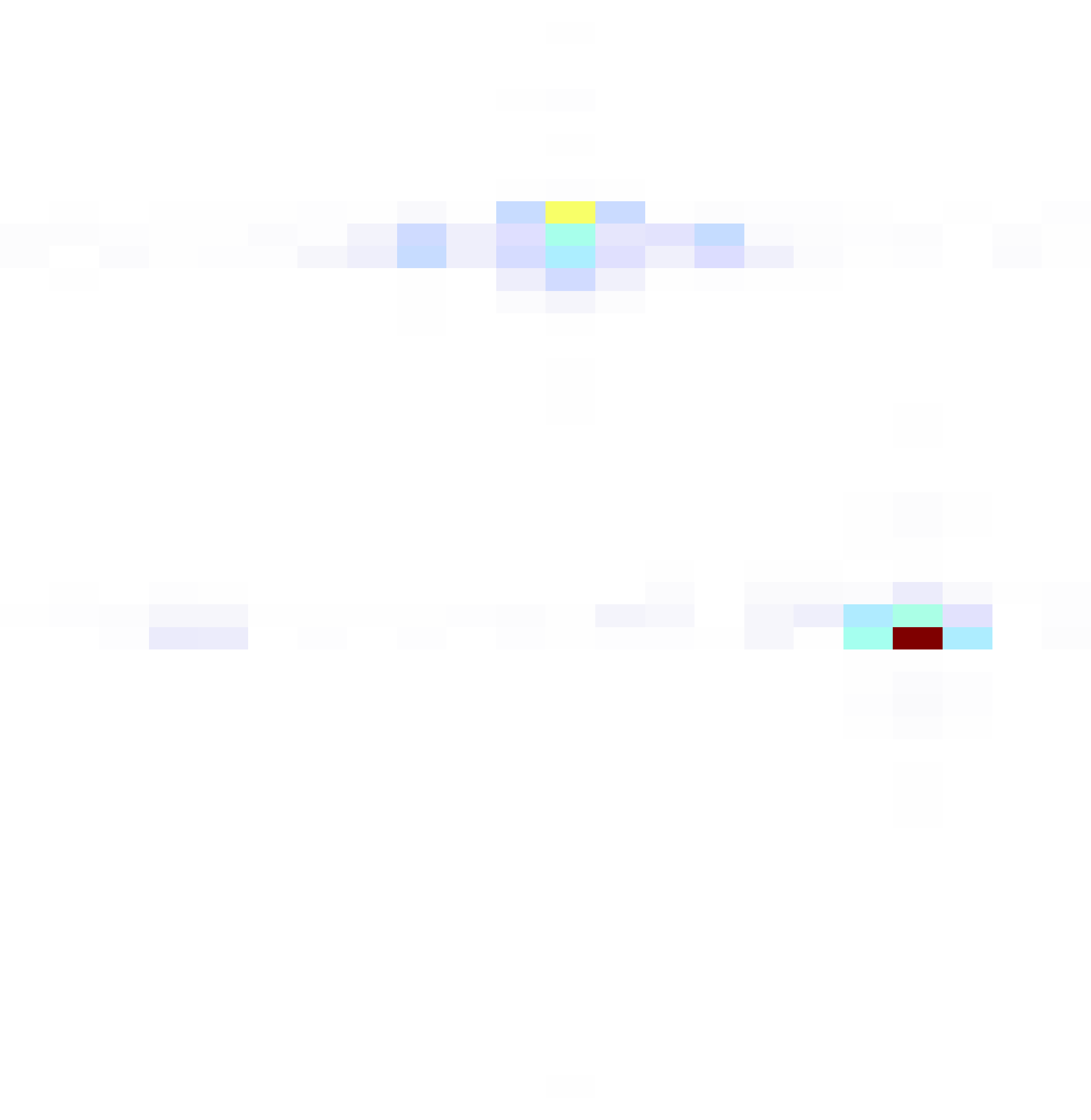};

      	\addplot[
      		mark=o,
                only marks,
      		mark size=15pt,
      		color=blue] 
      		table[x=speed,y=range] {
                speed range
                0 23.8
                };

      	\addplot[
      		mark=o,
      		only marks,
      		mark size=12pt,
      		color=red] 
      		table[x=speed,y=range] {
                speed range
                2 13.3
                };
            
        \node[draw=none, color=blue] at (-0.5, 28) {Clutter};
        \node[draw=none, color=red] at (1.5, 16.8) {Target};

    \end{axis}

\end{tikzpicture}

%% file: Figures/scrap_before_update_2.tikz
\begin{tikzpicture}

    \begin{axis}[
        axis on top,
        width=0.42\textwidth,
        %scale only axis,
        %axis equal=true,
        grid=major,
        enlargelimits=false,
        xmin=-2.5, xmax=2.5,
        xtick={-2, ..., 2},
        xlabel near ticks,
        ymin=5, ymax=30,
        ytick={0,5,...,30},
        ylabel near ticks,
        xlabel={Rel. Velocity  [m/s]},
        ylabel={Range [m]},
        label style={font=\footnotesize},
        tick label style={font=\footnotesize},
        legend style={font=\footnotesize},
        legend pos = north east,
        legend style=
        	{fill=white, 
        	fill opacity=0.4, 
        	draw opacity=1, 
        	text opacity=1, 
        	nodes={scale=1, transform shape}, 
            /tikz/every even column/.append style={column sep=0.1cm}
        	},
        %colorbar horizontal,
        colorbar,
        point meta min=0,
        point meta max=3074.07,
        colormap name=jet_inue,
        colorbar style=
            {ylabel={Power}, 
            at={(1.05, 0)},
            anchor = south,
            width = 0.3cm,
            ytick={500, 3000},
            ylabel style={yshift=0.8cm},
            every axis/.append style=
                {font=\footnotesize}
            }
        ]

      \addplot[forget plot] graphics[xmin=-3.3, xmax=3, ymin=0, ymax=30] {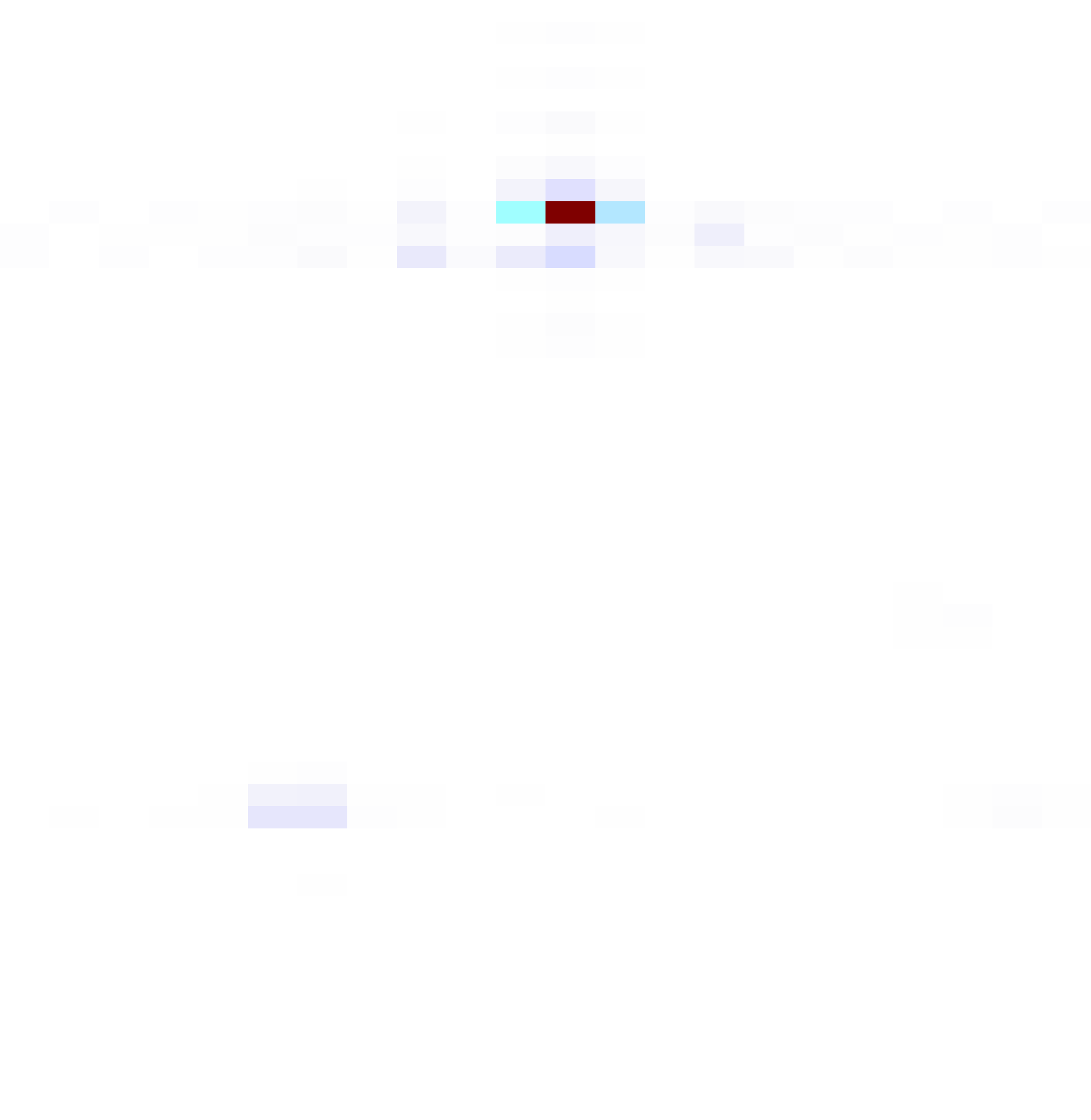};

      	\addplot[
      		mark=o,
                only marks,
      		mark size=15pt,
      		color=blue] 
      		table[x=speed,y=range] {
                speed range
                0 23.8
                };

      	\addplot[
      		mark=o,
      		only marks,
      		mark size=12pt,
      		color=red] 
      		table[x=speed,y=range] {
                speed range
                -1.57 8.2
                };

        \node[draw=none, color=blue] at (-0.5, 28) {Clutter};
        \node[draw=none, color=red] at (-1.05, 11.5) {Target};

    \end{axis}

\end{tikzpicture}

%% file: Figures/scrap_after_update_2.tikz
\begin{tikzpicture}

    \begin{axis}[
        axis on top,
        width=0.42\textwidth,
        %scale only axis,
        %axis equal=true,
        grid=major,
        enlargelimits=false,
        xmin=-2.5, xmax=2.5,
        xtick={-2, ..., 2},
        xlabel near ticks,
        ymin=5, ymax=30,
        ytick={0,5,...,30},
        ylabel near ticks,
        xlabel={Rel. Velocity  [m/s]},
        ylabel={Range [m]},
        label style={font=\footnotesize},
        tick label style={font=\footnotesize},
        legend style={font=\footnotesize},
        legend pos = north east,
        legend style=
        	{fill=white, 
        	fill opacity=0.4, 
        	draw opacity=1, 
        	text opacity=1, 
        	nodes={scale=1, transform shape}, 
            /tikz/every even column/.append style={column sep=0.1cm}
        	},
        %colorbar horizontal,
        colorbar,
        point meta min=0,
        point meta max=168.62,
        colormap name=jet_inue,
        colorbar style=
            {ylabel={Power}, 
            at={(1.05, 0)},
            anchor = south,
            width = 0.3cm,
            ytick={20, 160},
            ylabel style={yshift=0.8cm},
            every axis/.append style=
                {font=\footnotesize}
            }
        ]

      \addplot[forget plot] graphics[xmin=-3.3, xmax=3, ymin=0, ymax=30] {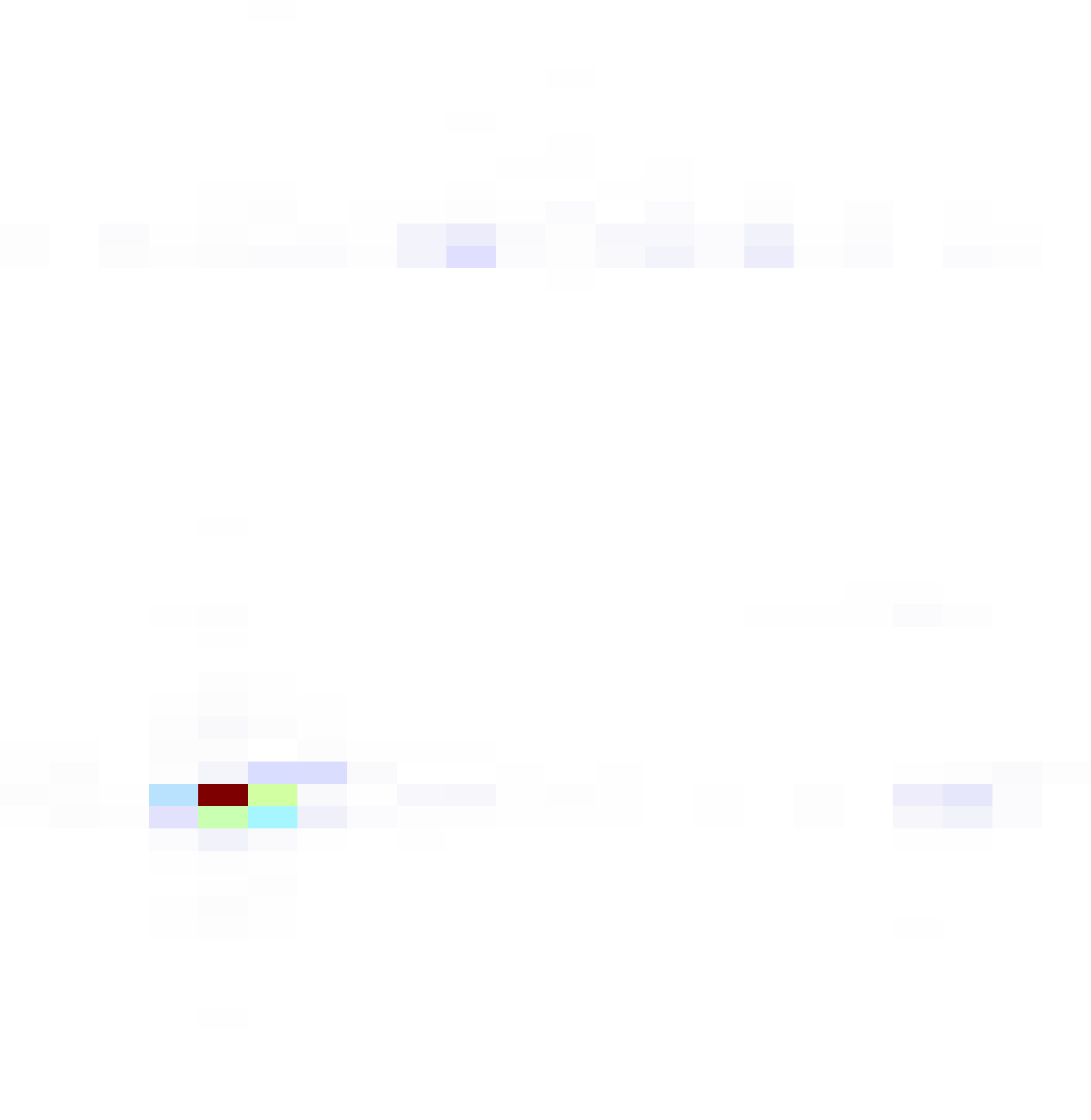};

      	\addplot[
      		mark=o,
                only marks,
      		mark size=15pt,
      		color=blue] 
      		table[x=speed,y=range] {
                speed range
                0 23.8
                };

      	\addplot[
      		mark=o,
      		only marks,
      		mark size=12pt,
      		color=red] 
      		table[x=speed,y=range] {
                speed range
                -2 8.2
                };

        \node[draw=none, color=blue] at (-0.5, 28) {Clutter};
        \node[draw=none, color=red] at (-1.5, 11.5) {Target};

    \end{axis}

\end{tikzpicture}

%% file: Content/Conclusion.tex
\section{Conclusion}\label{sec:conclusion}
% \todo[inline]{SM: rephrased here and there, adding a bit more emphasis on the numerical benefits and including outdoor as scenarios of interest. Feel free to revert of course
% \newline MH: It's fine, thanks (although I'm not a fan of the ``one order of magnitude"; to me it's imprecise.)}
In this work, we have introduced \gls{scrap}, which enables clutter removal that can adapt to dynamic clutter environments. This is done by twisting the concept of exponential smoothing to track the clutter multi-dimensional components in the \gls{csi} matrix. To do this efficiently, we properly scale new measurements and stack them with the previous clutter components scaled by their singular values in a unique matrix. The meaningful clutter components are determined using our proposed threshold based on the \gls{mp} distribution.     

Simulation results demonstrated that \gls{scrap} better isolates the sensing target from residual clutter than \gls{crap}, allowing to achieve one order of magnitude lower missed detection probabilities at low noise power. Moreover, we validated the proposed approach with measurements using the \gls{isac} \gls{poc} in the ARENA2036, highlighting \gls{scrap}'s benefits in a real-world environment with a strong dynamic clutter component.

In the future, we plan to conduct extensive measurements to examine the impact of the smoothing parameter and other parametrization options in different scenarios, \egc outdoors.

%% file: Content/Acknowledgment.tex
\section*{Acknowledgments}
The authors would like to thank Rolf Fuchs, Alexander Felix, and Thorsten Wild for their support during the development of this work.

This work was developed within the KOMSENS-6G project, partly funded by the German Ministry of Education and Research under grant 16KISK112K.